\documentclass[sigconf]{acmart}
\usepackage{xcolor}
\usepackage{enumitem}
\usepackage{subfigure}
\usepackage{booktabs} 
\usepackage{graphicx}
\usepackage{amsmath}
\usepackage{siunitx}
\usepackage{gensymb}
\usepackage{color}
\usepackage{soul}
\usepackage{multirow}
\usepackage{float}            
\usepackage{algorithm}        
\usepackage{algorithmic}      
\AtBeginDocument{%
  }

\copyrightyear{2025}
\acmYear{2025}
\setcopyright{acmlicensed}\acmConference[UIST '25]{The 38th Annual ACM Symposium on User Interface Software and Technology}{September 28-October 1, 2025}{Busan, Republic of Korea}
\acmBooktitle{The 38th Annual ACM Symposium on User Interface Software and Technology (UIST '25), September 28-October 1, 2025, Busan, Republic of Korea}
\acmDOI{10.1145/3746059.3747767}
\acmISBN{979-8-4007-2037-6/2025/09}

\begin{document}

\title{Wrist2Finger: Sensing Fingertip Force for Force-Aware Hand Interaction with a Ring-Watch Wearable}

\author{Yingjing Xiao}
\authornote{Both authors contributed equally to this research.}
\affiliation{%
  \institution{East China Normal University}
  \city{Shanghai}
  \country{China}}
\email{51275901140@stu.ecnu.edu.cn}

\author{Zhichao Huang}
\authornotemark[1]
\affiliation{%
  \institution{East China Normal University}
  \city{Shanghai}
  \country{China}}
\email{51275901117@stu.ecnu.edu.cn}

\author{Junbin Ren}
\affiliation{%
  \institution{East China Normal University}
  \city{Shanghai}
  \country{China}}
\email{51265901135@stu.ecnu.edu.cn}

\author{Yuting Bai}
\affiliation{%
  \institution{South China University of Technology}
  \city{Guangzhou}
  \country{China}}
\email{202321055492@mail.scut.edu.cn}

\author{Haichuan Song}
\affiliation{%
  \institution{East China Normal University}
  \city{Shanghai}
  \country{China}}
\email{hcsong@cs.ecnu.edu.cn}

\author{Zhanpeng Jin}
\affiliation{%
  \institution{South China University of Technology}
  \city{Guangzhou}
  \country{China}}
\email{zjin@scut.edu.cn}

\author{Yang Gao}
\authornote{Yang Gao is the corresponding author}
\affiliation{%
  \institution{East China Normal University}
  \city{Shanghai}
  \country{China}}
\email{gaoyang@cs.ecnu.edu.cn}




\begin{teaserfigure}
\centering
\includegraphics[width=0.95\textwidth]{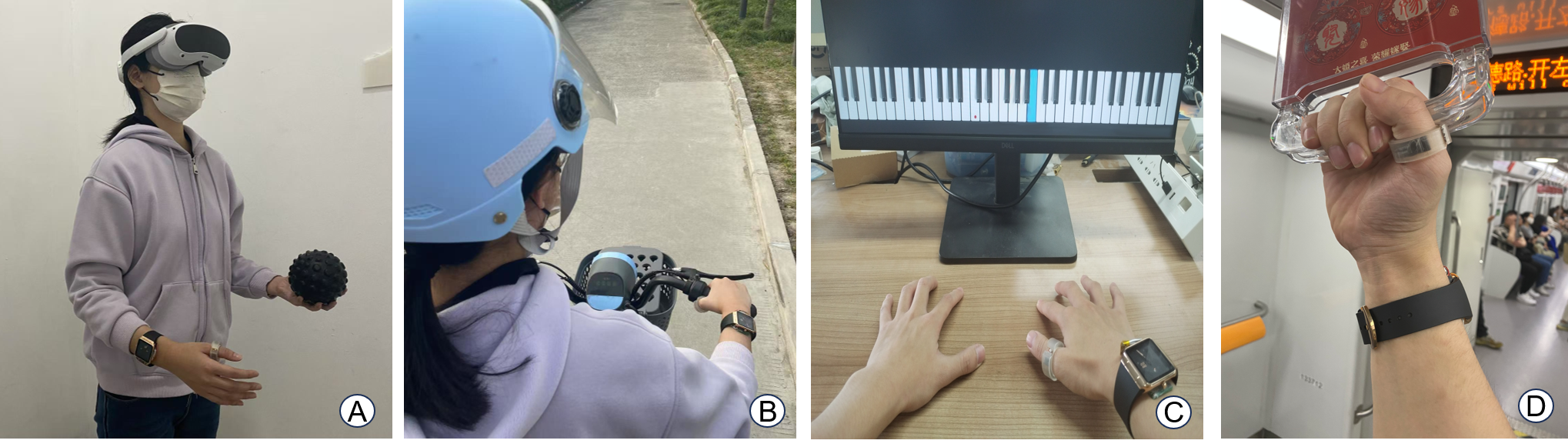}
    \caption{Daily Interactions with Wrist2Finger. Wrist2Finger enables force-aware interactions, even when the hands are occupied. (A) VR interaction with a daily object, (B) cycling while maintaining a firm grip, (C) playing a virtual piano with expressive fingertip control, and (D) holding onto subway handrails. }
    \label{fig:application_scenario}
\end{teaserfigure}

\begin{abstract}
Hand pose tracking is essential for advancing applications in human-computer interaction. Current approaches, such as vision-based systems and wearable devices, face limitations in portability, usability, and practicality. We present a novel wearable system that reconstructs 3D hand pose and estimates per-finger forces using a minimal ring-watch sensor setup. A ring worn on the finger integrates an inertial measurement unit (IMU) to capture finger motion, while a smartwatch-based single-channel electromyography (EMG) sensor on the wrist detects muscle activations. By leveraging the complementary strengths of motion sensing and muscle signals, our approach achieves accurate hand pose tracking and grip force estimation in a compact wearable form factor. We develop a dual-branch transformer network that fuses IMU and EMG data with cross-modal attention to predict finger joint positions and forces simultaneously. A custom loss function imposes kinematic constraints for smooth force variation and realistic force saturation. Evaluation with 20 participants performing daily object interaction gestures demonstrates an average Mean Per Joint Position Error (MPJPE) of 0.57 cm and a fingertip force estimation (RMSE: 0.213, r=0.76). We showcase our system in a real-time Unity application, enabling virtual hand interactions that respond to user-applied forces. This minimal, force-aware tracking system has broad implications for VR/AR, assistive prosthetics, and ergonomic monitoring.

\end{abstract}

\maketitle

\renewcommand{\shortauthors}{Xiao et al.}

\section{Introduction and Background}

Accurate hand tracking is foundational to human-computer interaction (HCI), powering applications from virtual and augmented reality (VR/AR) \cite{signLanguage, Du2022AMG} to prosthetics control and sign language interpretation \cite{Wu2022FullFiberAY, Khomami2021PersianSL}. Despite significant progress in capturing precise hand gestures, a critical dimension remains largely unexplored—force-aware interactions, or the capability to sense and interpret the user's exerted fingertip forces during interactions. Existing interaction techniques that lack force sensing severely limit the realism, expressiveness, and effectiveness of HCI applications. For example, distinguishing gentle touches from firm grasps can drastically improve the immersion and realism of VR experiences, enable nuanced gesture controls, and prevent potential harm in rehabilitation scenarios by monitoring and regulating the user's grip force.

However, integrating reliable force sensing into everyday wearable systems remains a major challenge. Vision-based solutions, including depth cameras \cite{MajdoubBhiri2023HandGR, leapmotion} and RGB cameras \cite{9903078, 9628050}, offer high spatial accuracy but require fixed infrastructures, face occlusion issues, raise privacy concerns, and constrain user mobility. Wearable alternatives, such as smart gloves \cite{2021Hybrid, Dong2021DynamicHG} and multi-sensor wristbands or armbands \cite{10623717, Parate2014RisQRS, Karheily2022sEMGTF}, offer mobility but introduce comfort, bulkiness, and complexity constraints. Even minimalist ring-based solutions (e.g., OmniRing \cite{zhou2023one}) typically require multiple rings or specialized sensing modalities (e.g., acoustic sensing \cite{yu2024ring}), hindering practical daily usability.

Ideally, a wearable solution should be socially acceptable, unobtrusive, and capable of capturing rich hand dynamics—especially the subtle yet crucial dimension of finger forces. Single-ring wearables offer promising minimalism but inherently lack the capability to fully sense complex, multi-finger movements or applied force. Existing studies attempt to address this limitation either by deploying additional custom hardware (e.g., sonar-based RingaPose, bio-impedance-based Z-Pose) or by requiring multiple rings and wrist-worn units \cite{zhou2023one}, both of which compromise usability. These challenges motivate our key research question:
\textbf{Can we achieve accurate, rich, and force-aware hand interactions using only a single ring and minimal complementary sensing?}

\begin{figure}[!tbp]
\centering
\includegraphics[width=0.4\textwidth]{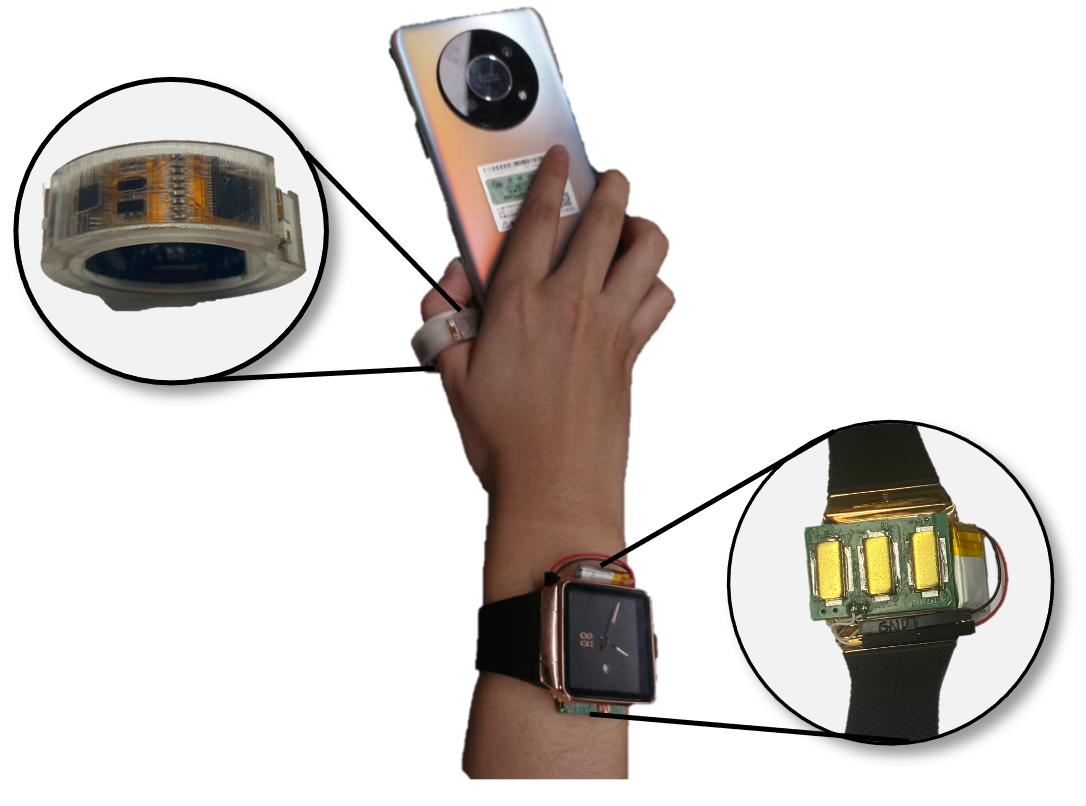}
\vspace{-10pt}
\caption{Ring-watch Hardware.}
\label{fig:finger-wrist-rationale}
\vspace{-15pt}
\end{figure}

In response, we present Wrist2Finger, a novel ring-watch wearable system designed to deliver force-aware interactions in an ergonomic, practical, and everyday wearable setup. As shown in Figure~\ref{fig:finger-wrist-rationale}, Wrist2Finger combines two complementary sensors: (1) a single IMU-equipped ring worn on one finger to provide local kinematic context, and (2) a single-channel electromyography (EMG) sensor embedded in a wrist-worn device, capturing muscle activations from forearm muscles controlling all fingers. Critically, EMG not only enriches the tracking of fingers not equipped with sensors but also serves as a reliable physiological proxy for fingertip force estimation, given the direct relationship between muscle activation and finger exertion.

Our approach leverages multimodal sensing to unlock both comprehensive hand pose reconstruction and fine-grained fingertip force estimation. Unlike purely kinematic trackers, Wrist2Finger captures the user's internal muscle effort, enabling a deeper and more intuitive interaction dimension. This capability opens new opportunities for applications such as immersive VR experiences, expressive gesture-based controls, or assistive technologies that monitor grip safety and intensity.

The primary contributions of this paper are as follows:
\begin{enumerate} \item A Minimalist, Force-Aware Wearable System: Wrist2Finger is the first wearable system combining a single IMU-equipped ring and a smartwatch-integrated single-channel EMG sensor, significantly improving portability and user comfort over traditional gloves or multi-ring setups.
\item Cross Attention-based Multimodal Fusion Model: We design a dual-branch transformer network employing cross-modal multi-head attention for dynamic fusion of IMU and EMG data, substantially enhancing hand pose tracking and force estimation accuracy compared to single-modality or simple fusion approaches.
\end{enumerate}

By directly addressing limitations in current vision-based and wearable approaches, Wrist2Finger moves closer toward the vision of always-available, force-aware hand interactions. In the following sections, we detail our system's hardware and algorithm design, report extensive evaluations and a VR demonstration scenario, and discuss implications for future wearable interactions. The code and hardware designs are publicly available at: \url{https://github.com/wood-3545/Wrist2Finger}.

\section{Related Work}
\begin{table*}[ht]
\centering
\caption{Wrist2Finger and other hand-based interaction method. "-" = Not Reported}
\begin{tabular}{|p{3.0cm}|p{2.8cm}|p{2.5cm}|p{1.8cm}|p{3.0cm}|p{2.5cm}|}
\hline
\textbf{Method} & \textbf{Form Factor} & \textbf{Sensing Modality} & \textbf{Body Part} & \textbf{Pose Estimation} & \textbf{Force Estimation} \\
\hline
CyberGlove \cite{mao2023simultaneous} & Glove & Flex Sensors, IMU & Hand & Continuous & -- \\
\hline
Ring-a-Pose \cite{yu2024ring} & Ring & IMU & Finger & Continuous & -- \\
\hline
DualRing \cite{liang2021dualring} & Ring & IMU & Finger & Continuous & -- \\
\hline
OmniRing \cite{zhou2023one} & Ring & IMU & Finger & Continuous & -- \\
\hline
picoRing \cite{takahashi2024picoring} & Ring & CFS
 & Finger & Discrete & -- \\
\hline

NeuroPose \cite{liu2021neuropose} & Armband & EMG & Wrist & Continuous & -- \\
\hline
Connan et al. \cite{connan2016assessment} & Armband & EMG & Wrist & -- & \checkmark \\
\hline
Vasconez et al. \cite{vasconez2022hand} & Armband & IMU + EMG & Hand & Discrete/continuous & -- \\
\hline
Jiang et al. \cite{jiang2017feasibility} & Wristband & IMU + EMG & Wrist & Discrete/continuous & -- \\
\hline

Our Method & Ring and Wristband & IMU + EMG & Finger & Discrete/continuous & \checkmark \\
\hline
\end{tabular}
\label{tab:method_comparison}
\end{table*}

\subsection{Wearable Hand Pose Tracking Approaches}
Hand pose tracking technology has long been explored for applications ranging from virtual and augmented reality interfaces to robotics and medical rehabilitation. Visual gesture tracking methods face significant limitations in complex environments, such as lighting changes, varying viewpoints, and occlusion, which make it difficult to guarantee accuracy and real-time performance\cite{mueller2019real,mueller2017real,yuan2017bighand2,zhang20163d}. Wearable solutions, on the other hand, offer a better solution to these issues, providing higher accuracy and reliability while reducing physical invasiveness, making them more feasible for everyday use. Current wearable solutions typically rely on instrumented gloves, IMU arrays, or EMG-based sensors. Data gloves, such as the CyberGlove, embed multiple sensors (e.g., flex sensors, multiple IMUs) to directly measure finger joint angles. While these systems are accurate, glove-based systems are bulky, intrusive, and unsuitable for long-term daily use due to comfort and practicality concerns. Recent minimalist solutions, such as ring-based trackers (e.g., Ring-a-Pose\cite{yu2024ring}, DualRing\cite{liang2021dualring}, OmniRing\cite{zhou2023one}), aim to overcome these limitations by placing miniature IMUs or alternative sensors (e.g., acoustic, impedance)\cite{siddiqui2020multimodal}  on individual fingers. However, these approaches either still require multiple rings or supplementary devices to achieve comprehensive pose estimation for all fingers, sacrificing some usability for accuracy. 

In contrast, our work adopts a minimalist wearable solution—a single IMU-equipped ring combined with an EMG sensor worn on the wrist—to balance usability, accuracy, and practical feasibility. By combining the IMU and EMG sensors, we are able to overcome the limitations of using a single sensor for complex gesture tracking. The EMG sensor provides muscle activation information from the fingers, reflecting fine-grained movement, while the IMU captures spatial dynamics to help track gesture posture changes. For instance, the study\cite{vasconez2022hand} demonstrates that combining EMG and IMU can achieve high classification accuracy. Additionally, NeuroPose\cite{liu2021neuropose} showcases the potential of EMG in 3D hand pose tracking, but without IMU integration, it limits the utilization of spatial dynamic information. Our approach, by combining both sensors, significantly reduces hardware complexity and physical invasiveness, while maintaining high accuracy, making it suitable for everyday use.

\subsection{Finger Force Estimation}
While extensive research exists on hand pose estimation, explicit finger force estimation remains less explored, particularly within compact wearable form factors. Existing finger force measurement\cite{zatsiorsky2000enslaving,meng2024unsupervised,zheng2024prediction,cho2022real} typically relies on instrumented gloves with embedded force sensors such as strain gauges or piezoresistive FlexiForce sensors placed directly at fingertips. For instance, gloves instrumented with FlexiForce sensors have been utilized to quantify precise fingertip forces during activities of daily living or rehabilitation tasks. However, these gloves are bulky and intrusive, limiting their practicality for everyday or consumer-oriented VR/AR interactions\cite{mao2023simultaneous}.

Another indirect approach is to infer fingertip forces from surface EMG signals measured from forearm muscles. Previous work\cite{jiang2017feasibility} demonstrated EMG signals correlate well with overall grip force due to muscular activation intensity; however, they typically rely on multi-channel electrode arrays (such as the Myo armband) to achieve accurate force estimation. Multi-channel EMG approaches\cite{you2010finger,connan2016assessment} provide richer spatial resolution, distinguishing muscular activity patterns for individual fingers but significantly increasing hardware complexity, power consumption, and sensitivity to electrode placement, thus decreasing user comfort and practicality for long-term wear\cite{bangaru2020data}.

In contrast, our approach leverages a minimal single-channel EMG sensor at the wrist and complements it with finger-level spatial context from the ring-mounted IMU. By using advanced neural network architectures (dual-branch transformers with attention-based fusion), we effectively map limited EMG input to precise multi-finger force estimations, overcoming the limitations of single-modality solutions while significantly reducing hardware complexity. Our method bridges the gap between practicality (single-channel EMG) and detailed fingertip force inference, thereby filling a critical gap in wearable sensing research.

\subsection{Multimodal Sensor Fusion for Wearables}
Recent research highlights the benefits of multimodal sensing strategies, combining multiple complementary sensor modalities to enhance robustness, accuracy, and functional range of hand tracking systems\cite{tanweer2019development, georgi2015fusion,siddiqui2020multimodal,lopes2017hand,Khomami2021PersianSL}. Systems fusing IMU sensors with surface EMG sensors have been successfully demonstrated for gesture classification, significantly outperforming single-modality setups due to the complementary nature of motion and muscle activation data. For example, Khomami et al. demonstrated improved gesture classification accuracy by fusing forearm-mounted EMG and IMU data, showing EMG's strength in capturing muscular effort and IMU's efficacy in capturing spatial and dynamic motion characteristics.

Despite these successes, prior multimodal systems usually required multiple sensors (e.g., full EMG armbands with multiple electrodes combined with multiple IMUs or sensors on the hand or forearm), which inevitably compromises ease of wear and user convenience. Moreover, existing multimodal systems have primarily focused on classifying discrete gestures or gross pose estimation, neglecting fine-grained continuous pose reconstruction and finger-specific force estimation.

Our proposed solution addresses these limitations by adopting a minimal multimodal configuration: one ring-mounted IMU and one wrist-mounted EMG sensor. By employing a transformer-based dual-branch architecture with multi-head cross-modal attention, we dynamically fuse spatial (IMU) and physiological (EMG) features, enabling precise hand pose reconstruction and continuous fingertip force estimation simultaneously. We further improve realism by incorporating biomechanical and physiological constraints in training (smoothness, saturation, regularization). This architecture uniquely enables both detailed multi-finger pose reconstruction and continuous force estimation with minimal instrumentation, extending the state-of-the-art significantly toward practical, everyday use.

\section{Sensing Integration of Finger and Wrist}

\begin{figure}[!tbp]
    \centering    \includegraphics[width=0.49\textwidth,height=0.17\textheight]{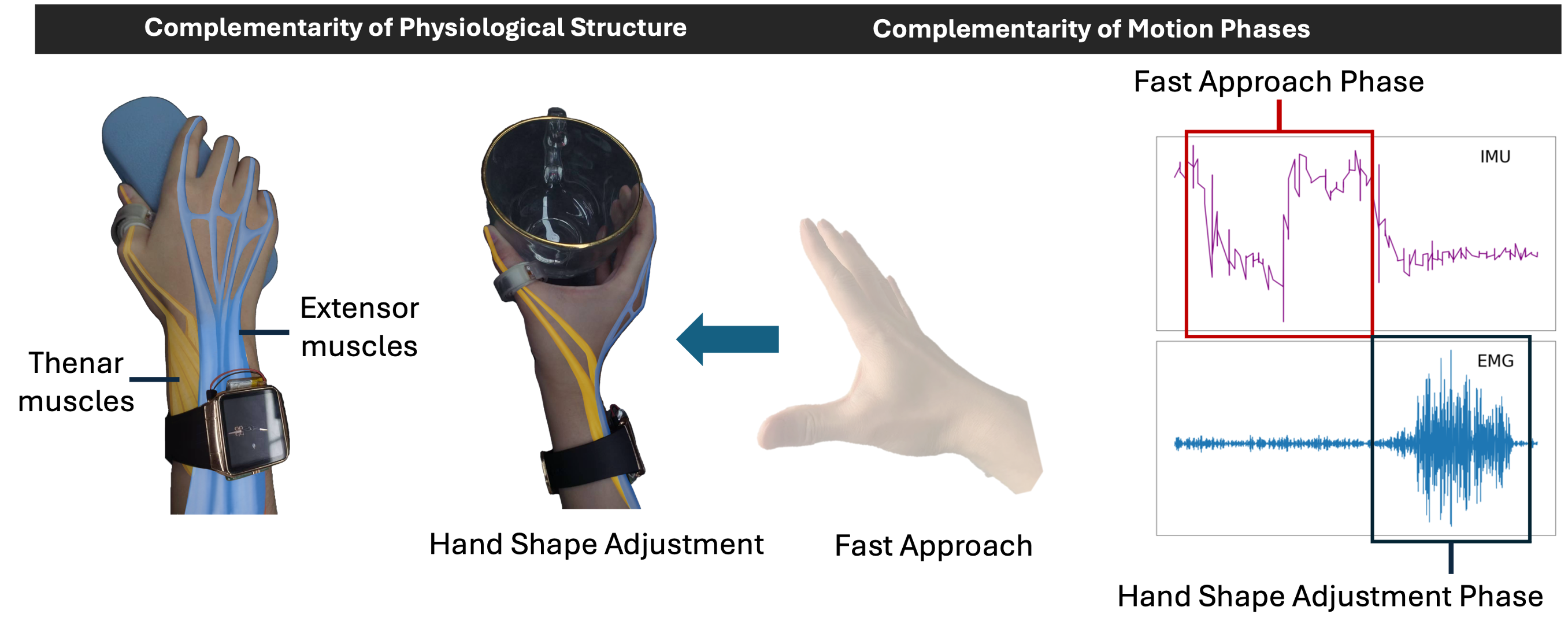}
    \caption{Complementary sensing of hand grasp interaction from both physiological and motion phase.}
    \label{fig:complementary_sensing}
    \vspace{-15pt}
\end{figure}

\subsection{Complementary Sensing}
Hand anatomy reveals a distinct biomechanical separation between the muscle groups controlling the thumb and those responsible for the other four fingers. The thumb is primarily governed by the thenar muscles, including the abductor pollicis brevis, flexor pollicis brevis, and opponens pollicis, while the extensor muscles, such as the extensor digitorum communis and extensor indicis, control the movement of the index, middle, ring, and little fingers. These muscle groups are both anatomically and functionally distinct, highlighting a natural separation that can be leveraged for precise hand interaction.

Our system integrates data from two complementary wearable sensors: a single-channel electromyography (EMG) sensor embedded in the dorsal side of a smartwatch and an inertial measurement unit (IMU) in a compact ring worn on the thumb. This dual-sensing approach capitalizes on the complementary capabilities of EMG and IMU: the EMG sensor captures muscle activation signals from the wrist-adjacent muscle groups responsible for finger movements \cite{lieber1992architecture}, while the IMU tracks the kinematic motion of the thumb with high precision.

As illustrated in Figure \ref{fig:complementary_sensing}, these two sensors work in harmony, reflecting the natural biomechanical independence of the thumb and the other fingers. The IMU provides insight into the dynamic kinematic phases of thumb motion, while the EMG sensor is sensitive to muscle activation patterns, ensuring robust hand pose tracking even during complex, high-speed gestures. The complementarity of these two sensing modalities is particularly effective during different motion phases, such as when subtle changes in thumb posture or force exertion are coupled with corresponding shifts in muscle activation, enabling more accurate force-aware hand interaction.

Unlike previous ring-based methods that require multiple rings on different fingers for accurate hand tracking \cite{DBLP, ZHOU2022LearningOT, zhou2023one, inproceedings}, our approach simplifies the hardware setup by using only a single ring worn on the thumb. Additionally, while other EMG-based systems often rely on multi-channel arrays to achieve higher resolution \cite{article, 9956792, 10623717, Karheily2022sEMGTF}, our single-channel EMG solution maintains a high level of performance without the complexity of embedding multiple sensors into a smartwatch form factor. This enables seamless integration into everyday wearables while ensuring effective hand interaction tracking.

\begin{figure*}[!tbp]
\centering
\includegraphics[width=0.9\textwidth,height=0.3\textheight]{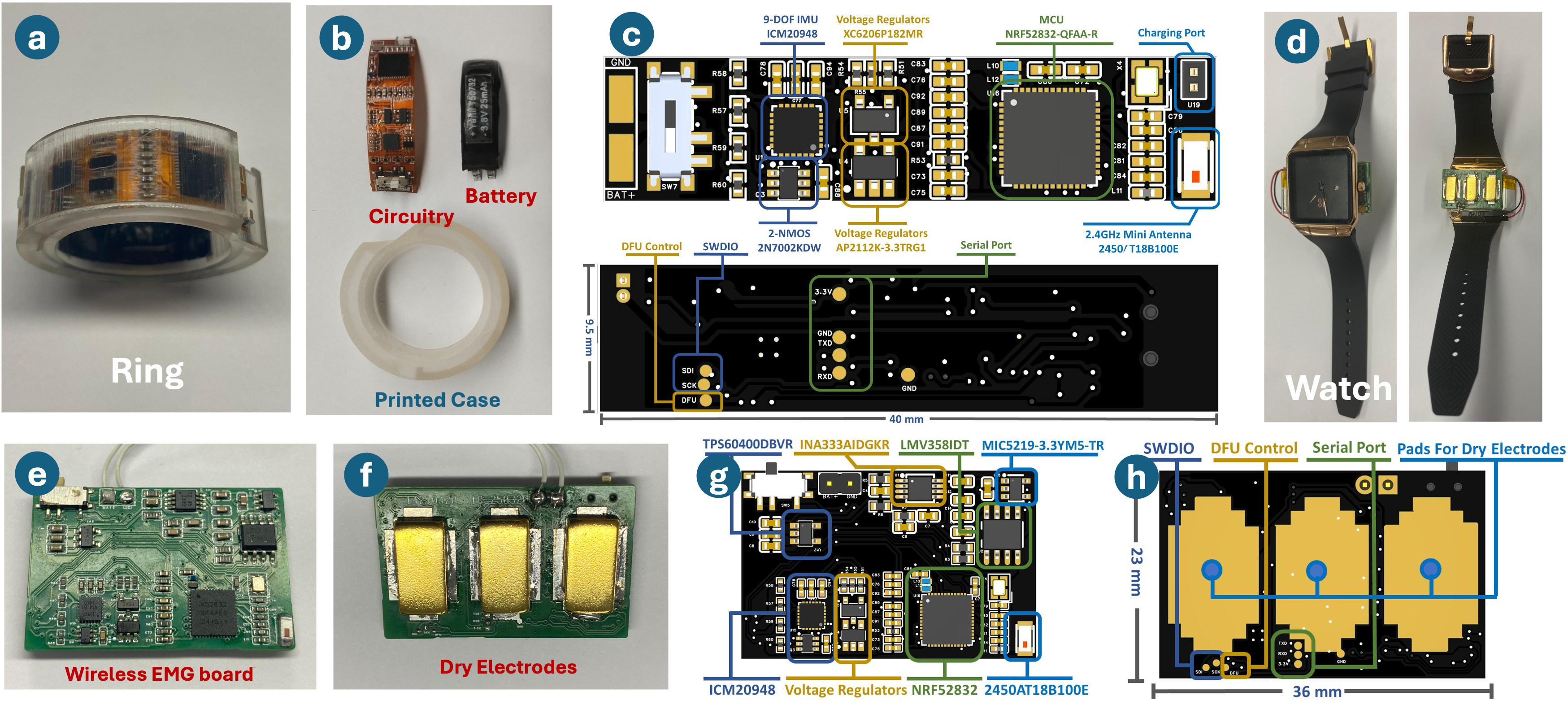}
\vspace{-10pt}
\caption{Ring-Watch Wearables. (a) Ring (b) main components of the ring prototype. (c) circuitry of the ring. (d) sensor-side of the watch prototype. (e) wireless EMG process board (front-side). (f) gold-plated EMG dry electrodes (back-side). (g) circuitry of the EMG sensor. (h) layout of EMG electrodes}
\label{fig:hardware_intro}
\vspace{-15pt}
\end{figure*}

\subsection{Design Consideration}

The specific placement and configuration of sensors in Wrist2Finger were carefully chosen based on considerations of biomechanics, sensor complementarity, and practical usability. The decision to place the IMU sensor on a single ring primarily worn on the thumb or index finger is driven by the need for capturing precise local motion data, as these fingers often exhibit the most diverse and frequently utilized movements during hand interactions. The ring-mounted IMU efficiently tracks spatial and orientation information of the finger, providing detailed kinematic context essential for accurately reconstructing hand poses.

Conversely, the decision to embed a single-channel EMG sensor into a smartwatch worn on the wrist leverages the physiological characteristic that finger force exertion is predominantly generated by forearm muscles. The EMG sensor captures muscle activation patterns, effectively reflecting user intent and the exerted fingertip forces. By placing this sensor on the wrist, the system benefits from a stable and socially acceptable wearable form factor that minimizes obtrusiveness and maximizes user comfort for prolonged everyday use.

\subsection{Wearables Design}

\subsubsection{Ring Prototype} As shown in Fig.\ref{fig:hardware_intro}, the ring prototype consists of a flexible circuitry, an arc-shaped battery, and a 3D-printed case, designed for both compactness and wearability. The prototype is built around the NRF52832 MCU\cite{nrf52832Ref}, which integrates a 2.4 GHz BLE radio transceiver and an ARM Cortex-M4 32-bit processor in a compact 6mm x 6mm package. The IMU, ICM20948\cite{ICM20948Ref}, provides 9-axis data and communicates with the MCU via I2C, chosen for its simplicity and efficiency. We use a XC6206P182MR voltage regulator to provide 1.8V for IMU and a AP2112K-3.3TRG1 voltage regulator to provide 3.3V for board. A 2N7002KDW integrating two independent NMOS tubes is used to handle 1.8V and 3.3V connections. To optimize Bluetooth data transmission, the rate is set at 55Hz with a payload of 64 bytes per data package. The companion application has been redesigned to integrate functionalities from the Nordic OTA app\cite{NordicAppRef} and the Adafruit app\cite{AndroidAppRef,IosAppRef}. 

Compared to existing smart rings designed for continuous pose tracking (e.g., Ring-a-Pose\cite{yu2024ring}: 148mW, ssLOTR\cite{zhou2022learning}: 198mW, and PeriSense\cite{wilhelm2020ring}: 474mW), our ring prototype prioritizes real-time IMU data transmission with a streamlined architecture. By including only essential peripheral circuits, it achieves a more compact design (9.5 mm width, 40 mm length) and lower power consumption (56 mW). In full-featured streaming mode, with real-time data collection and high-rate Bluetooth transmission, the ring operates at a current of 13 mA, enabling continuous operation for over 2 hour on a 25 mAh arc Li-ion battery.





\subsubsection{Integreating EMG sensors in a watch} 

To collect electromyographic (sEMG) data, we designed a dry electrode sEMG printed circuit board (PCB) integrated into a watch for close contact and ease of use (Fig. \ref{fig:hardware_intro}). The compact PCB (23 mm × 36 mm) combines circuits for sEMG signal acquisition, processing, and wireless transmission, along with an IMU (ICM20948) for motion posture calibration. It adds only 8.8 g and fits comfortably within the 42–51 mm diameter range of standard smartwatch enclosures.

The PCB has three gold-plated copper electrodes arranged in parallel to align with arm muscles, capturing differential voltage generated by muscle activity. This signal is amplified by the INA333AIDGKR instrumentation amplifier, which features high input impedance, programmable gain, and high common-mode rejection, effectively amplifying weak EMG signals while rejecting noise. The amplified signal is filtered by the LMV358IDT op-amp, suppressing out-of-band noise and ensuring compatibility with the ADC.

To support dual-rail operation, a TPS60400DBVR charge-pump inverter provides a $-3.3\,\mathrm{V}$ supply, while the MIC5219-3.3YM5-TR LDO regulator ensures a stable 3.3 V output, reducing power-supply noise. The system uses an NRF52832 MCU, which integrates a 2.4 GHz BLE transceiver, to receive both EMG and IMU data and transmit them to the upstream device.

Despite integrating sEMG signal acquisition, processing, wireless transmission, and an additional IMU, our system can still operate efficiently at a current as low as approximately 13\,mA, with a BLE data rate of 55\,Hz and 64 bytes per data packet. Powered by a 100\,mAh high-efficiency lithium battery, the device can run for over 7 hours on a single charge, corresponding to a power consumption of about 13\,mAh per hour---only around 3\% of a typical 425\,mAh smartwatch battery.

To ensure good wearability, we use biocompatible, gold-plated electrodes. In our short-term wear tests (up to two hours), no skin irritation or discomfort was reported by participants.

\subsubsection{Calibration}
Before use, we perform a short calibration. The IMU is calibrated for bias and orientation offset on the user’s finger (instructing the user to hold their hand flat and relax). The EMG channel is calibrated by recording a brief rest signal (no hand activity to set a baseline) and a maximum voluntary contraction (user makes a very tight fist for a couple of seconds to gauge the EMG amplitude range). These allow us to normalize EMG readings per user. With these calibrations, the system is ready to capture data for the deep learning model.

\subsubsection{Data Capture}
The ring transmits IMU data (accelerometer, gyro readings) to the watch via BLE in real-time. The watch collects its own EMG data and timestamps both streams, then forwards them to a paired smartphone or PC for processing. In our current prototype, the heavy processing (neural network inference) runs on a laptop, but in principle a modern smartwatch or phone could handle it. The slight latency introduced by wireless streaming is around 20–30 ms, which is negligible for our application.

\section{Multimodal Hand Force Estimation}

\begin{figure}[!tbp]
    \centering    \includegraphics[width=0.49\textwidth,height=0.17\textheight]{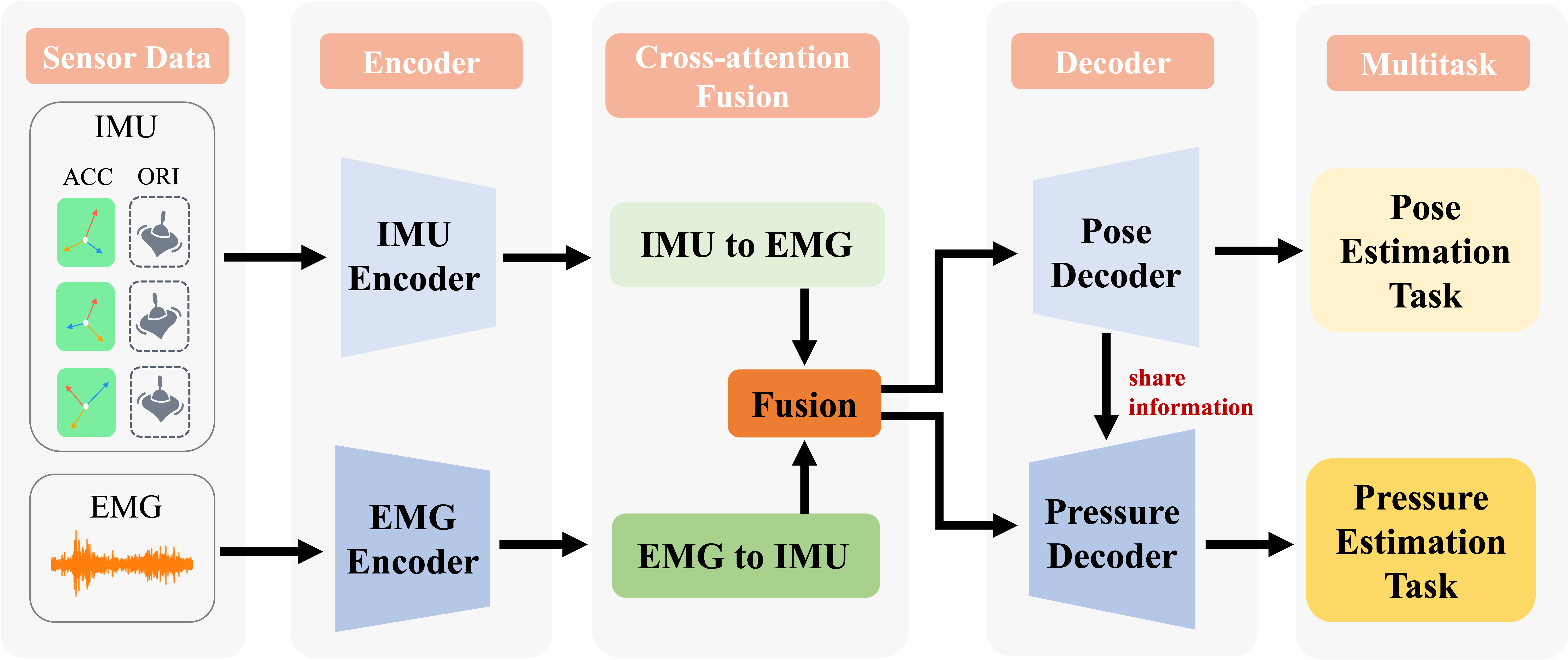}
    \caption{The architecture of the hand pose and pressure reconstruction network that integrates IMU and EMG data. }
    \label{fig:framework}
    \vspace{-10pt}
\end{figure}

To map the multimodal sensor data to hand pose and force estimates, we designed a dual-branch deep neural network with an encoder-decoder based architecture, as illustrated in Figure~\ref{fig:framework}. The model takes as input time-synchronized sequences from the IMU and EMG sensors and outputs the predicted hand pose (joint angles or 3D joint positions) and forces at the fingertips.

\subsection{Input Representation}

We first preprocess the raw sensor streams into features suitable for the model. The original IMU data is sampled at 100 Hz and is subsequently downsampled to 30 Hz to reduce redundancy and computational cost. For each window of 30 frames (corresponding to a 1-second interval at 30 Hz), the IMU feature at each frame consists of a 3-axis acceleration vector and a 9-dimensional rotation matrix from the ring and watch IMUs, resulting in a total IMU feature dimension of 24 per frame. The EMG data is synchronized and downsampled to the same rate, and the rectified and smoothed EMG amplitude (1 channel) is expanded to 6 dimensions to match the input requirements of the transformer. Therefore, each input window has 30 frames, with IMU features of dimension 24 and EMG features of dimension 6 per frame. The model outputs the 3D positions of 21 hand joints and force values for five fingertips for each window. Finally, positional encoding is applied to each time step for both IMU and EMG feature sequences, enabling the model to capture temporal dependencies.

\subsection{Dual-Branch Encoding}
The network has two parallel branches: one for IMU and one for EMG. The hidden feature dimension throughout the network is set to 128. Both the IMU and EMG branches are encoded using two-layer transformers (4 heads), with the IMU branch including a preliminary 1D convolution and the EMG branch using a two-layer LSTM. Cross-modal fusion uses bidirectional 4-head multi-head attention.

Each branch is a transformer encoder that processes the sequence of its respective modality. The IMU branch encoder is responsible for capturing temporal patterns and relationships in the motion data. It can learn, for example, the characteristic sequence of accelerations when a certain gesture (like a pinch or a grasp) is performed. The EMG branch encoder similarly captures temporal patterns in muscle activation—e.g., a gradual increase in EMG corresponding to finger tightening. Importantly, both encoders produce a sequence of latent embeddings for each time step in the window. We then employ a multi-head cross-attention mechanism to fuse the information from the two branches. This cross-attention mechanism adaptively aligns the two modalities per sample, allowing IMU queries to locate temporally shifted EMG activations (e.g., leading or lagging signals during force exertion), and vice versa, enabling EMG queries to pinpoint hand-pose frames corresponding to peak muscle engagement. In our design, we use the IMU branch outputs as “queries” and the EMG branch outputs as “keys/values” in a cross-modal attention layer. This allows the model to learn dynamic associations between motion and muscle signals. For instance, if a certain spike in the EMU signal at time t corresponds to a subtle motion captured by the IMU at time t, the attention mechanism will assign a higher weight linking those two features. Multi-head attention (we use 4 heads) lets the model consider different aspects of correlation (perhaps one head focuses on overall signal energy while another on temporal alignment offsets). By stacking a couple of these cross-attention layers, the network effectively merges the modalities into a joint representation. This approach is inspired by recent successes of transformers in multimodal learning, where attention can flexibly align features across modalities, though to our knowledge it has not been extensively applied to wearable sensor fusion for hand tracking before.

\subsection{Feature Fusion and Decoding}
The outputs of the attention layers are concatenated and passed through a series of feed-forward layers to produce a fused feature vector representing the current window’s hand state. This fused latent vector now contains information from both IMU and EMG pertinent to the hand’s configuration and applied forces. The final stage of the network is an output decoder implemented as a multi-layer perceptron (MLP). The decoder is designed to predict two sets of outputs: (1) the pose of the hand (we represent pose as the 3D positions of key hand joints or equivalently the joint angles – in our implementation we output 21 joint positions for the hand skeletal model), and (2) the forces at each of five fingertips. We structure the MLP such that it has two parallel output heads: one for pose and one for force. Each head is further structured to predict each finger’s values somewhat independently. Specifically, the force head has 5 units (one per finger) and the pose head has 5 groups of units corresponding to finger joint angles. By using a linear activation at the final layer (no coupling nonlinearity among outputs), we ensure a form of output linearity – meaning each finger’s predictions are a linear combination of the learned features, which helps avoid impossible inter-finger interactions in the output. In practice, this means the network can predict, say, a high force on the index finger and a low force on the middle finger without interference, reflecting that our sensing and model allow separate finger outputs. This design was chosen because early experiments with a fully-connected joint output tended to produce correlated errors (e.g., predicting all forces high or all fingers curled together). By decoupling outputs, we encourage the model to treat each finger’s state independently unless the data truly indicates correlation.

\subsubsection{Loss Function}

Training the network end-to-end on pose and force requires a carefully designed loss function. We incorporate several terms:

The fused features are passed through fully connected layers to predict the final 3D hand pose and force information. To optimize the sensor fusion process and enhance reconstruction accuracy, we design a weighted loss function that integrates contributions from IMU data, EMG signals, and joint angles. Each loss component is tailored to address specific aspects of hand pose reconstruction and force estimation, ensuring complementary strengths from each data modality.

The IMU loss captures spatial and kinematic information by calculating the difference between the predicted and ground truth hand poses. Additionally, it incorporates the discrepancy between predicted and actual force values, enhancing the model's ability to perceive hand-applied forces during dynamic interactions. Leveraging the high temporal resolution of IMU data, this loss term not only ensures precise tracking of hand motion but also improves the timeliness and accuracy of force estimation, making it especially effective for fast or complex dynamic motion scenarios.
The EMG loss evaluates the consistency between the predicted hand poses and the muscle activation patterns detected by the EMG sensors. It introduces intent-driven information, enabling the model to infer subtle finger and hand movements that may not be directly observable from kinematic data alone. In addition to pose estimation, this loss also computes the discrepancy between predicted and actual force values, leveraging the correlation between muscle activity and exerted force. This dual-role design allows the model to simultaneously enhance pose accuracy and force prediction based on neuromuscular intent.
The angle loss measures discrepancies in 3D joint angles, ensuring that the reconstructed hand poses adhere to anatomical constraints. This loss term not only improves the physical plausibility of predicted poses but also enhances robustness under noisy or incomplete data conditions.

By combining these three loss components, the model effectively balances spatial precision, muscle intent, and anatomical consistency, resulting in a robust and accurate hand pose reconstruction framework.

The total loss function is expressed as:
\[
L_{\text{total}} = \lambda_{\text{IMU}} L_{\text{IMU}} + \lambda_{\text{EMG}} L_{\text{EMG}} + \lambda_{\text{Angle}} L_{\text{Angle}}
\]
where \( \lambda_{\text{IMU}} \), \( \lambda_{\text{EMG}} \), and \( \lambda_{\text{Angle}} \) are weighting factors used to balance the influence of each loss component.


To output both joint position and force information, we add two separate output layers at the end of the model: a hand pose output layer and a force output layer. Specifically, the hand pose output layer is configured with a dimension of num\_joints * 3, where num\_joints = 21, corresponding to the 3D coordinates (x, y, z) of each joint. This layer decodes the encoded features into the spatial positions of the joints for 3D hand pose prediction.

Meanwhile, the force output layer is set with a dimension of num\_fingers, where num\_fingers = 5, representing the force values of the five fingers. This layer decodes the fused features into the individual finger force outputs, enabling precise prediction of applied pressure.

With these two output modules, the model can not only reconstruct high-accuracy 3D hand gestures but also simultaneously estimate the force exerted by each finger during interactions, providing a comprehensive understanding of both motion and force behaviors of the hand.

\section{Experiments}

\subsection{Data Collection}
\begin{figure}[!tbp]
    \centering    \includegraphics[width=0.42\textwidth]{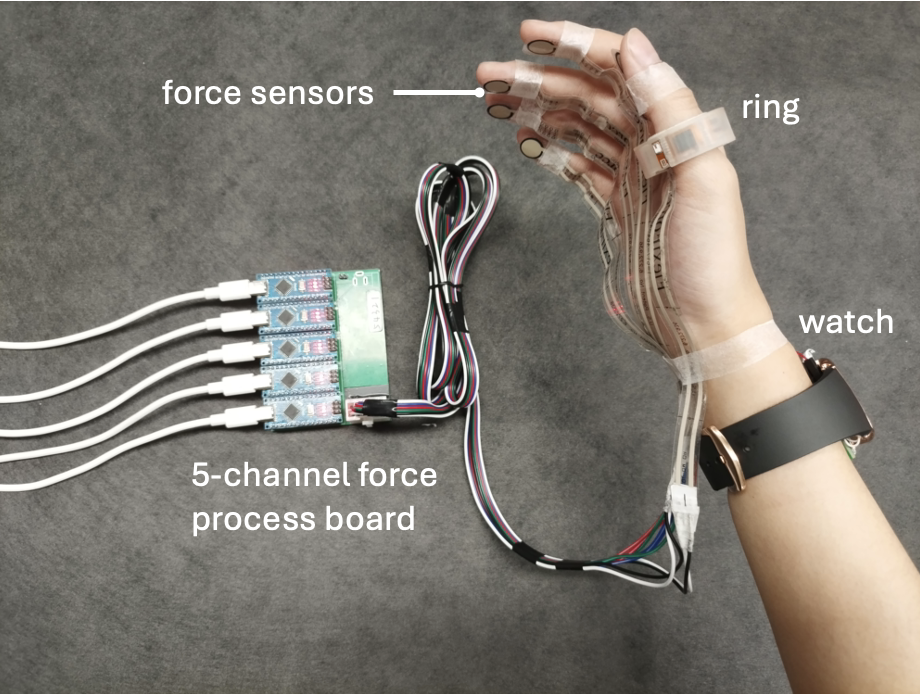}
    \caption{Experiment setup. Ground-truth forces were measured via a five-channel fingertip force sensor.}
    \label{fig:experiment_setup}
    \vspace{-10pt}
\end{figure}

\paragraph{Participants} We conducted a comprehensive evaluation of our ring-watch system with 20 participants. Each participant provided informed consent and was instructed on the use of the device. The study protocol was approved by our institutional review board. Participants wore the ring on their thumb finger and the smartwatch with EMG on their dominant hand’s wristas illustrated in Figure~\ref{fig:experiment_setup}.

\paragraph{Gestures and objects set} We defined a set of 20 representative hand gestures and grasp actions, inspired by activities of daily living, incorporating a wide variety of hand shapes, grip strengths (e.g., grasping, pinching, squeezing), and objects with diverse properties (e.g., weight, surface material, and shape). The gestures involved interacting with common objects to elicit natural hand poses and force profiles. Examples include: holding a coffee mug (wrap grasp with moderate force), gripping a screwdriver (precision pinch using thumb and index), squeezing a tennis ball (full-hand fist squeeze with high force), and lifting a small thermos (combined fingers grasp). We also included free-hand gestures without objects for variety, such as a thumbs-up, open hand stretch, fist clench, and pointing. The diversity of object interactions ensures a wide range of force outputs, from light touch (e.g., holding a tape) to strong force (e.g., squeezing a ball). These 20 gestures cover various hand configurations (pinch, power grasp, precision grip, etc.) and force levels, and are designed to interact with objects of varying weights, surface materials, and shapes, providing a broad range of realistic hand interactions.

\paragraph{Procedure} Each participant performed each gesture multiple times (5 repetitions) to provide sufficient data. During a session, participants stood or sat in front of a table with the required objects. An experimenter provided verbal instructions (e.g., “grasp the mug as if drinking”), and participants performed the action for about 10 seconds, followed by releasing or relaxing their hand. Natural motion and force were encouraged—such as squeezing the ball at their normal strength rather than maximally unless that was their natural behavior. The sequence of gestures was randomized to avoid order effects or fatigue bias.
To mitigate the risk of model overfitting and ensure reliable data across repeated sessions, we incorporated a re-worn condition within the same experimental session. Each participant performed the same gesture sequence multiple times, with sufficient rest between repetitions to minimize fatigue. The study has been approved by the Internal Review Board of East China Normal University.

\paragraph{Ground-Truth force measurement} To obtain ground truth for the finger forces, we instrumented a data glove with force sensors. Specifically, we attached FlexiForce piezoresistive force sensors at the tip of each finger. These are thin film sensors that change resistance under applied pressure. We calibrated each sensor using known weights before the experiment. The FlexiForce sensors\cite{flexiforce} output an analog value proportional to the force; during tasks, they measured the pressure each fingertip applied on the object or in free air (contact pressure for pinch grips, etc.). These sensors can reliably measure forces up to 25 N, which aligns with typical maximal fingertip forces reported for daily object grasp tasks\cite{li2013directional}. The FlexiForce readings were recorded at 100 Hz via an Arduino data logger, and we synchronized these readings with the ring/watch data by a shared timestamp (all devices connected to the same computer clock). The force data serves as ground truth labels for training and evaluating the force estimation part of our model.

\paragraph{Ground-Truth pose measurement} To ensure accuracy while minimizing the impact on the user's hand experience, instead of using a motion-capture system with reflective markers or gloves, following recent practices \cite{10.1145/3613904.3642663, 10.1145/3613904.3642910, 10.1145/3526113.3545665}, we used a single RGB camera-based setup. Specifically, we adopted MediaPipe\cite{mediapipe}, a well-established 3D hand pose estimation tool with 1-cm normalized mean absolute error (NMAE), to extract hand keypoints directly from video. This lightweight and non-intrusive setup avoids discomfort commonly associated with markers or gloves, preserving natural hand behavior. 
The video was recorded at 60 Hz using a high-resolution smartphone camera (Redmi Note 12, 48MP) under controlled conditions—close distance and a high-contrast background—to ensure stable detection.

\subsection{Dataset and Implementation}

\begin{figure*}[!tbp]
\centering
\includegraphics[width=0.9\textwidth,height=0.22\textheight]{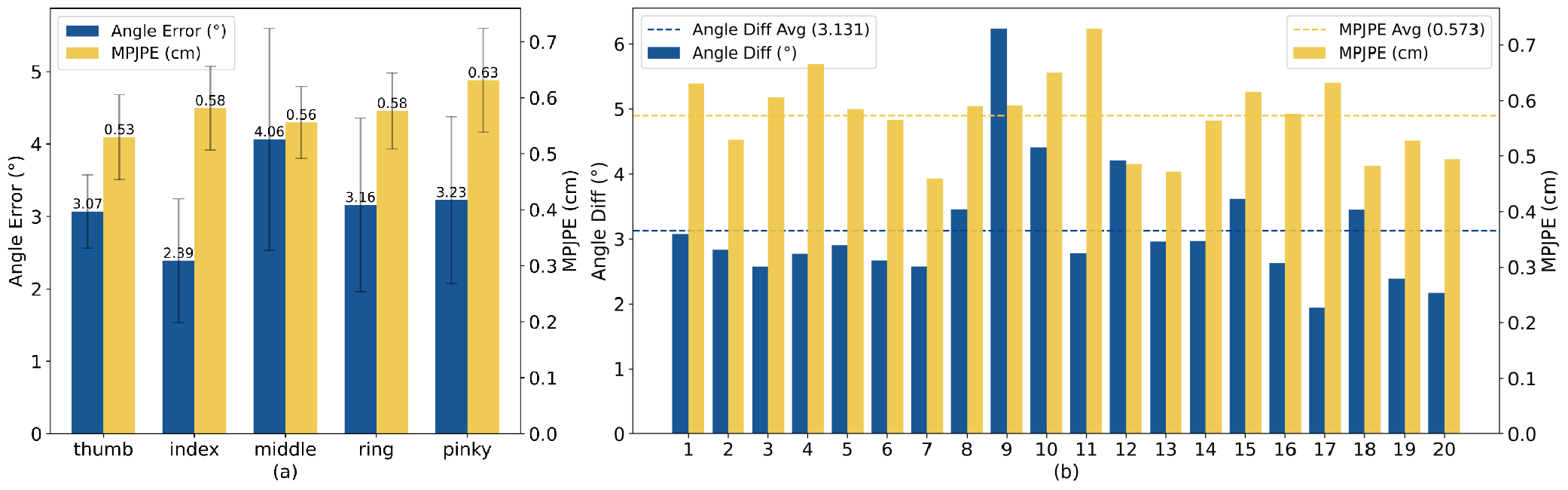}
\vspace{-5pt}
\caption{(a) Hand Pose Tracking Errors among different fingers; (b) Leave-One-User-Out Cross-Validation Results. The bar charts show the MPJPE (cm) and joint angle differences (°) for each participant, with the red dashed lines indicating the average values.} \label{fig:pose_performance}
\vspace{-5pt}
\end{figure*}

\begin{figure*}[!tbp]
\centering
\includegraphics[width=0.9\textwidth,height=0.22\textheight]{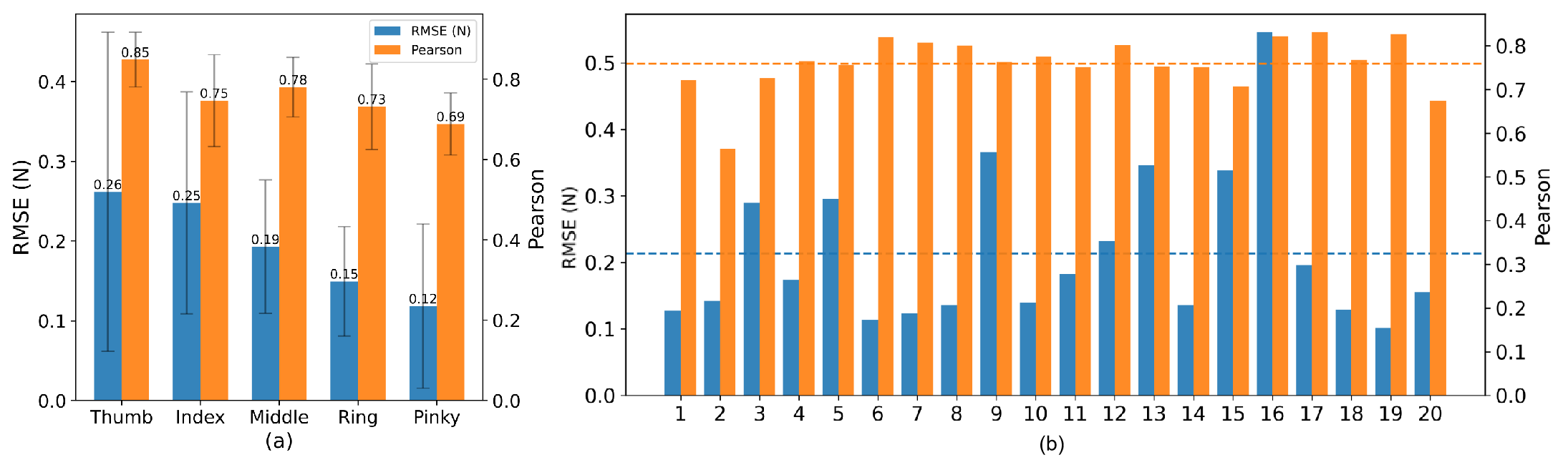}
\vspace{-10pt}
\caption{(a) Hand pressure estimation result among different fingers; (b) Leave-One-User-Out Cross-Validation results. The bar charts show the RMSE and Pearson for each participant, with the red dashed lines indicating the average values.} \label{fig:pressure_performance}
\vspace{-10pt}
\end{figure*}

\paragraph{Dataset preparation}: We segmented the continuous data into gesture trials. Each trial is labeled with the gesture type and associated object. We also annotated qualitatively the intended force level (e.g., low, medium, high) for analysis. In total, across 20 participants × 20 gestures × ~5 repetitions, our dataset contains approximately 2000 gesture trials, which amounts to about 2.5 hours of data per participant on average (some gestures are shorter, some longer). We split each participant’s data into training and testing segments. Additionally, for cross-user evaluation, we compiled a dataset that trains on 19 participants and tests on the held-out 20th.

\paragraph{Model Implementation:}  
Training uses a batch size of 256, Adam optimizer. All training is performed for 100 epochs per stage with early stopping. Pressure values are log-transformed and min-max normalized. Window size is 30, step size is 5.

The model was trained in multiple stages to achieve optimal performance. Initially, the model was trained end-to-end to establish a robust baseline, followed by fine-tuning specific components, such as the pose decoding layer and the EMG encoder, to enhance precision. Hyperparameters were carefully tuned to balance the contributions of the IMU, EMG, and angular loss components, with optimal values of $\lambda_{\text{IMU}} = 0.5$, $\lambda_{\text{EMG}} = 0.5$, and $\lambda_{\text{angle}} = 1.0$. This ensured a harmonious trade-off between modalities and effective minimization of angular discrepancies. All training was performed using the Adam optimizer with a batch size of 256 and a window size of 30 (step size 5). Additionally, learning rates of $0.001$ for training and $0.0001$ for fine-tuning, paired with a patience-based learning rate scheduler, facilitated stable convergence while avoiding overfitting. These adjustments collectively enabled the model to achieve precise and robust 3D hand pose estimation.

\paragraph{Evaluation Metrics:} The proposed framework's performance was assessed using established metrics, including Mean Per Joint Position Error (MPJPE) and joint angle differences. These metrics comprehensively evaluate the accuracy of 3D pose reconstruction, as well as the anatomical consistency of joint movements.

\section{Evaluation and Experimental Results}

We evaluate our system’s performance on two primary tasks: hand pose estimation and force estimation. We report quantitative accuracy metrics for each, and also examine the contribution of each component of our model via an ablation study. All results are reported on test data not seen during training.

\subsection{Pose Estimation Accuracy}

\subsubsection{Tracking Performance Among Different Fingers}
Fig.\ref{fig:pose_performance}(a) illustrates the tracking accuracy for each finger, with the thumb achieving the lowest MPJPE (0.625 cm) and joint angle difference (5.9°), while the little finger showed the highest errors (MPJPE: 0.935 cm, angle difference: 7.8°). The results highlight the impact of sensor proximity, with fingers closer to the IMU (e.g., thumb, index) exhibiting better performance. Anatomical variations and reduced motion range of smaller fingers, such as the little finger, contributed to increased errors. These findings underscore the need for sensor placement and algorithmic refinement to ensure consistent performance.


\subsubsection{Leave-One-User-Out cross-validation}

We conducted leave-one-user-out cross-validation to evaluate the generalizability of our framework. Figure \ref{fig:pose_performance}(b) illustrates the results for each participant, showing both the MPJPE and joint angle differences as bar charts. The blue and orange dashed lines represent the average performance across all participants for each metric.
The results demonstrate the model's robustness and accuracy across participants. MPJPE values ranged tightly from 0.520 cm to 0.934 cm, while joint angle differences remained consistently below 10°, with an average of 6.815°. This consistency underscores the framework's ability to generalize across diverse hand anatomies and gestures, likely due to the complementary integration of IMU and EMG data.

\subsubsection{Comparison with EMG-only Baseline}
We compared our system with NeuroPose \cite{liu2021neuropose}, a representative EMG-only method, on the PiMForce public benchmark dataset \cite{seo2024postureinformed}. Our approach achieved a joint angle error of 3.58°, closely matching NeuroPose's 3.41°, while using only a single EMG channel instead of a full 8-channel array. This demonstrates that our fusion of a ring-mounted IMU with minimal EMG sensing offers comparable pose accuracy at significantly lower hardware complexity.

\subsection{Force Estimation Accuracy}
For force estimation, we evaluate the predicted force at each fingertip against the ground truth from the FlexiForce sensors. 
We report the Mean Squared Error (MSE) in normalized force units, as well as the average absolute error in Newtons (after converting back).

Overall, the model achieved an average force MSE of 0.045 (in normalized 0–1 units), which corresponds to about ± 1.0 N error on average per finger (since our typical force range was 0–20 N). For context, a difference of 1 N is quite small – for example, the difference between holding a cup with 5 N vs 6 N is barely noticeable. The correlation between predicted and actual force time-series is high (average Pearson r = 0.76), indicating our model captures the trends of increasing/decreasing force well.


\paragraph{Examining per-finger force results} As shown in Fig.\ref{fig:pressure_performance}, the thumb and middle finger force estimates are the most accurate (average Pearson r > 0.8), which is expected since those fingers generate stronger EMG signals and often dominate grasps. The index and ring fingers have slightly higher error (r < 0.75). The little finger force had the highest error (r <0.7) and lower correlation, in part because the little finger’s contribution in many tasks is small and noisy relative to the sensor’s resolution. However, even for the little finger, the system can discern between “no force” and “some force” reliably – e.g., it won’t mistakenly output a high force when the little finger isn’t pressing at all (the false-positive rate for force above a small threshold was below 5\%). The force saturation constraint in training prevented wild errors; the maximum force our model ever predicted was ~1.1 in normalized units (which when converted was ~26 N, near the known physical max.

\paragraph{Comparing to different pressure estimation models} To assess the performance of our proposed Wrist2Finger model, we compared it against existing state-of-the-art finger pressure estimation networks that rely on multi-channel EMG data, which is typical of current wrist-worn solutions. Specifically, we evaluated three notable models: DCNN, LSTM, and CNN-LSTM, each representing different strategies for estimating finger pressure using EMG signals. As shown in Table\ref{tab:baseline}, our model outperforms all others in all metrics. Wrist2Finger achieves the lowest RMSE (0.213) and MAE (0.076), and the highest Pearson correlation (0.759), indicating advantage of accuracy in predicting finger pressure. This improvement is primarily due to the integration of a cross-channel attention strategy, which captures finger motion information from IMU data and enhances pressure estimation compared to EMG-only models.

\begin{table}[htbp]
\centering
\caption{Performance comparison with different pressure estimation models}
\label{tab:baseline}
\begin{tabular}{lccc}
\hline
\textbf{Method}              & \textbf{RMSE↓}  & \textbf{MAE↓}   & \textbf{Pearson↑} \\ \hline
DCNN \cite{su2021deep}                          & 0.255          & 0.118          & 0.230            \\
LSTM \cite{olsson2021end}                         & 0.266          & 0.111          & 0.155            \\
CNN-LSTM \cite{wahid2024semg}                     & 0.246          & 0.099          & 0.340            \\
\textbf{Wrist2Finger (Ours)} & \textbf{0.213} & \textbf{0.076} & \textbf{0.759}   \\
\hline
\end{tabular}
\end{table}


\subsection{Multi-modal Complementary Analysis among Interaction Gestures} 

\begin{figure}[!tbp] 
\centering \includegraphics[width=0.48\textwidth,height=0.24\textheight]{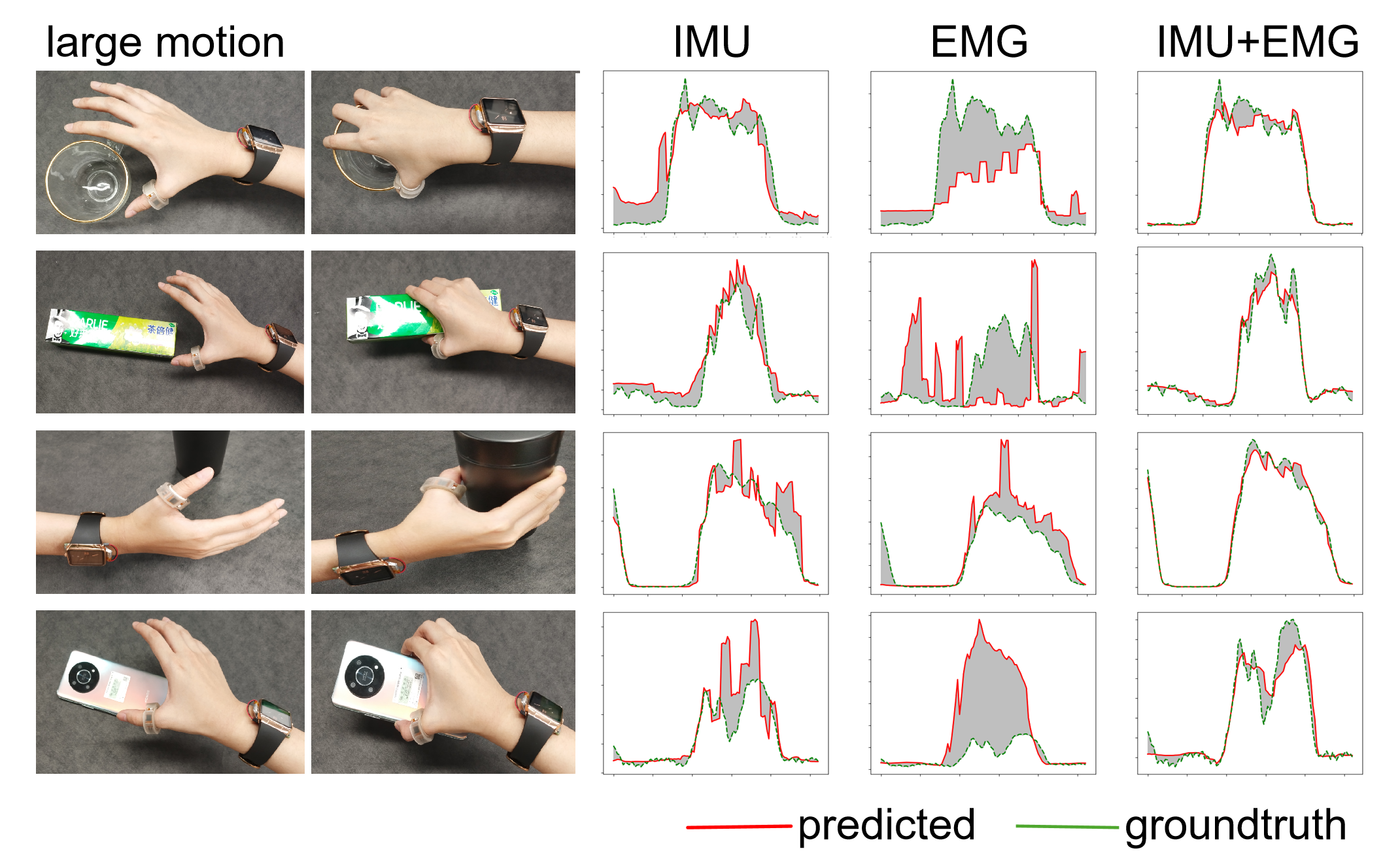}
\vspace{-10pt}
\caption{Modality contribution in pressure estimation under large motions. The right side shows the comparison between predicted and ground-truth thumb finger pressure under different input modalities. (pressure values normalized between 0 and 1.)}
\label{fig:IMU_EMG_error} 
\vspace{-10pt}
\end{figure}

\begin{figure}[!tbp] 
\centering \includegraphics[width=0.48\textwidth,height=0.24\textheight]{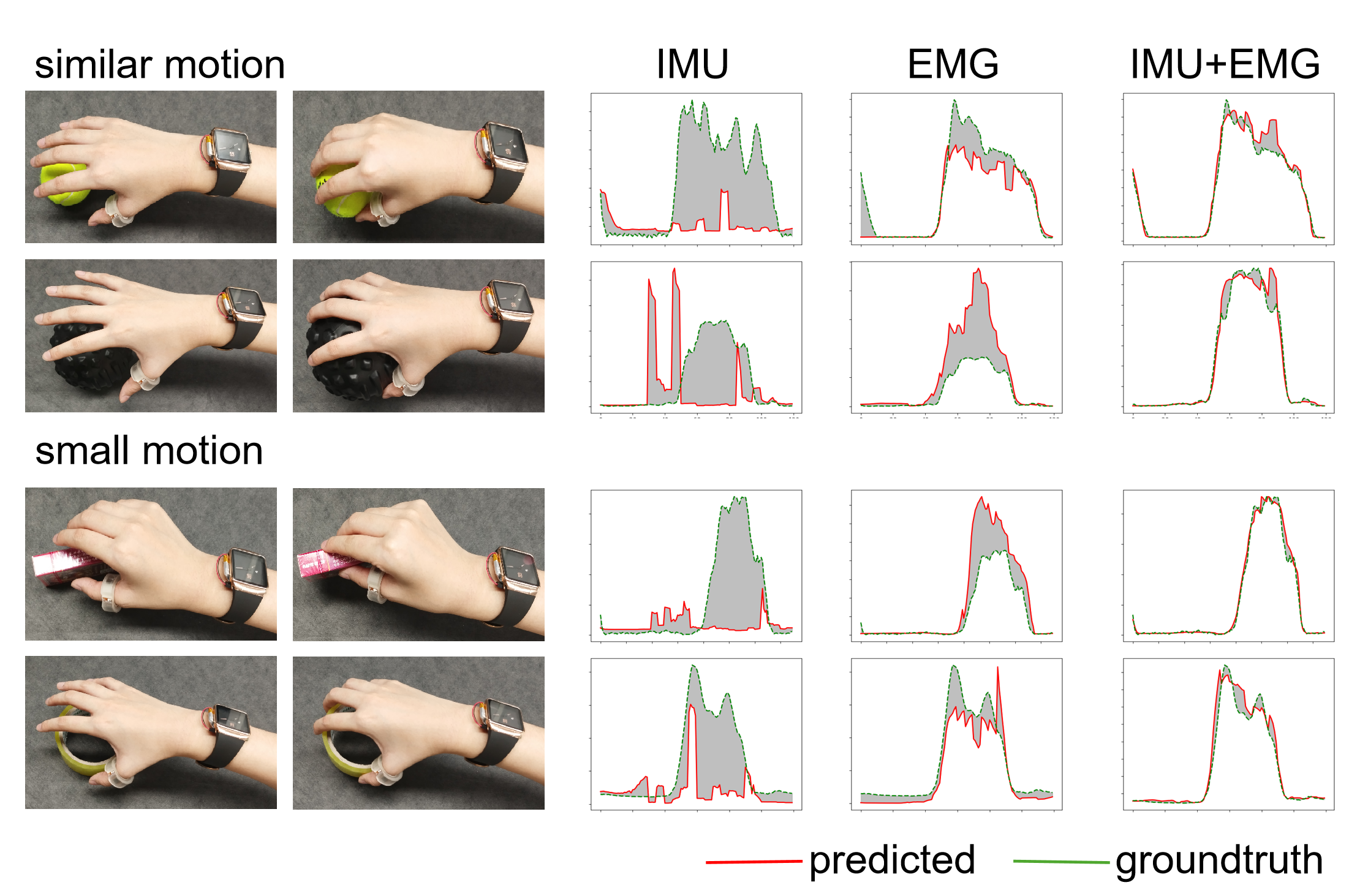}
\vspace{-10pt}
\caption{Modality contribution in pressure estimation under small and similar motions. The right side shows the comparison between predicted and ground-truth thumb finger pressure under different input modalities. (pressure values normalized between 0 and 1.)}
\label{fig:EMG_IMU_error} 
\vspace{-10pt}
\end{figure}

In this section, we explore the complementary contributions of the IMU and EMG modalities in estimating finger pressure during various hand interaction gestures. As shown in Figure \ref{fig:IMU_EMG_error} and \ref{fig:EMG_IMU_error}, the performance of the two modalities varies significantly depending on the motion characteristics of the hand gestures.

For gestures involving larger hand movements, such as those in the "large motion" category (e.g., holding a bottle or a glass), the IMU modality outperforms EMG alone. In these scenarios, the IMU-based single-modal prediction provides better accuracy in estimating pressure, as the large motion of the hand generates stronger and more easily detectable motion signals. This is particularly evident in the comparison of the IMU and EMG predictions, where the IMU predictions (in red) closely follow the ground truth (in green) for these gestures.

On the other hand, for gestures involving smaller motions (e.g., holding a small object like a tube or a tape), the EMG modality provides a stronger contribution. These gestures produce less noticeable movement, which makes it difficult for the IMU to accurately capture the nuances of finger motion changes. In these cases, the EMG modality, which is sensitive to muscle activity, provides a more reliable estimate of force changes. 

Additionally, for "similar motion" gestures (such as the consistent grip on different size objects), both IMU and EMG perform reasonably well. However, the complementary nature of the two modalities becomes more evident when combining IMU and EMG data, as shown in the "IMU + EMG" column, where the fusion of both modalities leads to predictions that more closely align with the ground truth. This highlights the advantage of multi-modal systems in handling a wider range of hand gestures, improving force estimation accuracy across various motion types.

In summary, IMU is particularly useful for larger, dynamic hand movements, while EMG excels in scenarios with smaller or more subtle finger gestures. The complementary nature of IMU and EMG underscores the importance of multi-modal sensing systems in providing accurate, robust force estimation across diverse hand interactions.

\subsection{Impact of Different Wearing Positions}
In the design of wearable systems for force-aware interaction, selecting the optimal position for sensor placement is crucial for maximizing performance. To this end, we evaluated the impact of different finger-wearing positions on the accuracy of pressure prediction, with the goal of identifying the most effective position for capturing force-related data. Our results indicate a notable performance variation across different fingers, with the thumb position yielding the best results.

\begin{figure}[!tbp] 
\centering \includegraphics[width=0.42\textwidth,height=0.24\textheight]{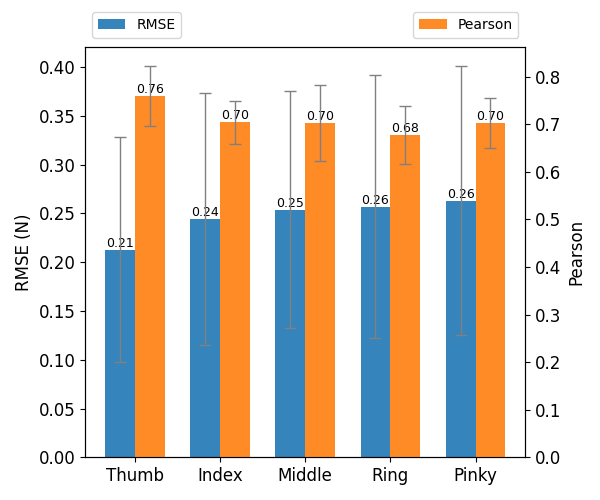}
\vspace{-10pt}
\caption{Pressure reconstruction performance across different wearing positions of the ring.}
\label{fig:error_per_finger} 
\vspace{-10pt}
\end{figure}

Specifically, the thumb demonstrated the lowest RMSE (0.21 N) and the highest Pearson correlation (0.76), outperforming all other fingers, as illustrated in Figure~\ref{fig:error_per_finger}. This suggests that the thumb's motion provides richer complementary information when combined with wrist EMG signals, facilitating more accurate pressure prediction. In contrast, while the index, middle, ring, and pinky fingers also show strong performance, their RMSE values and correlation scores are slightly lower, with the thumb's superior performance remaining evident.

These findings underscore the importance of strategically selecting the wearing position to optimize force estimation accuracy. In particular, the thumb's unique kinematics and its synergy with wrist-based muscle signals make it the most effective choice for force-aware interaction applications. However, our results also suggest that the system remains effective when worn on other fingers, providing flexibility in real-world applications.

\subsection{Impact of Anthropometric Data}
\begin{figure}[!tbp] 
\centering \includegraphics[width=0.48\textwidth,height=0.24\textheight]{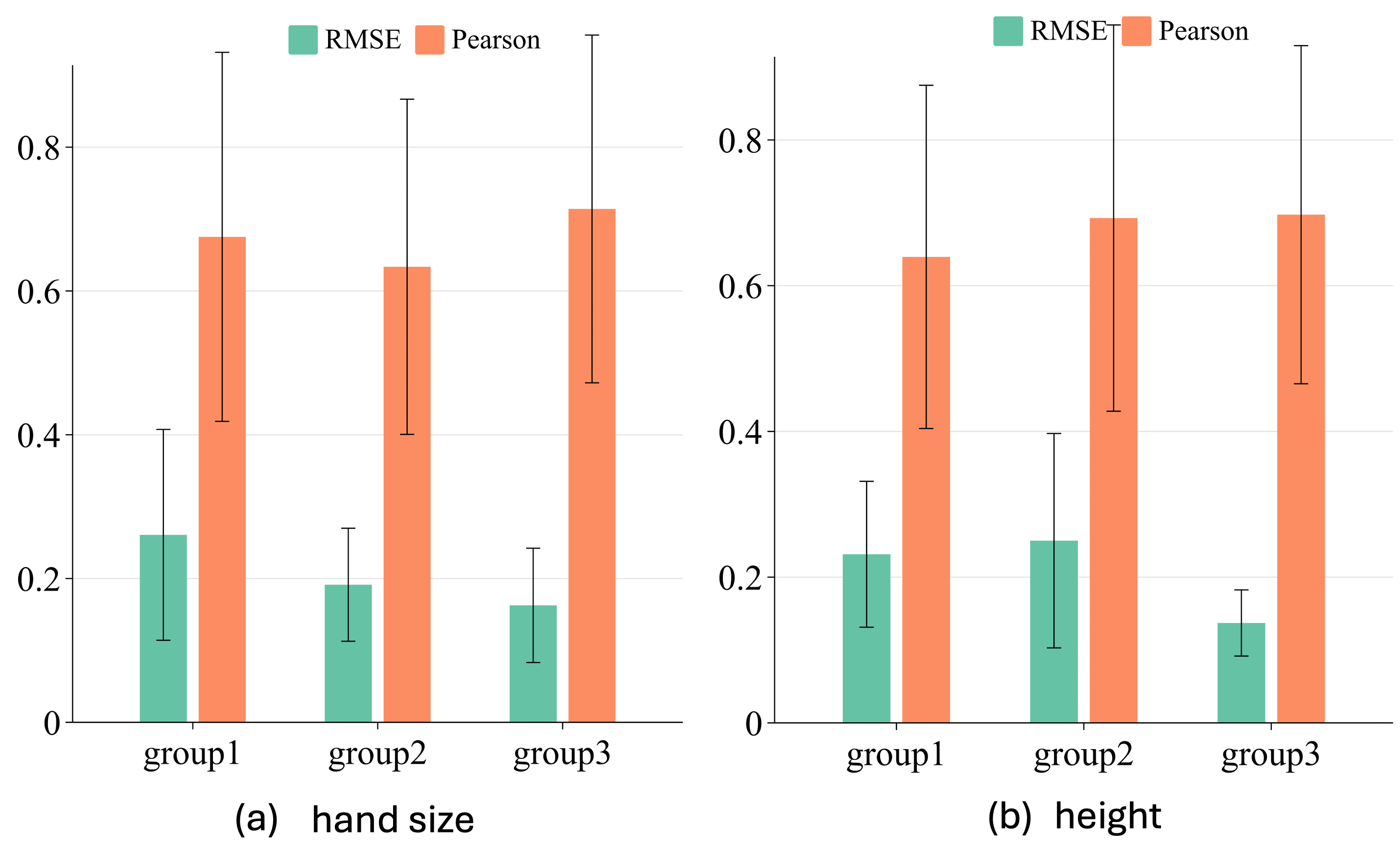}
\vspace{-15pt}
\caption{Performance across different anthropometric data}.
\label{fig:demographic_group} 
\vspace{-10pt}
\end{figure}

\paragraph{Hand Size} To assess the impact of anthropometric data on finger force estimation, we classified participants based on hand length into three distinct groups: Group 1 (hand length < 17.1 cm), Group 2 (hand length 17.1–19.6 cm), and Group 3 (hand length > 19.6 cm). This classification is particularly important, as hand size is known to influence both force application and perception during interactions.
The results in Figure \ref{fig:demographic_group}(a) show that Group 3, with larger hands, had the lowest RMSE, indicating more accurate force estimation. In contrast, Group 1, with smaller hands, had the highest RMSE, reflecting more difficulty in force prediction. Pearson correlation values followed a similar trend. 

\paragraph{Body Height} Participants were also grouped based on body height: Group 1 (< 165 cm), Group 2 (165–172 cm), and Group 3 (> 172 cm). As shown in Figure \ref{fig:demographic_group}(b), Group 3 achieved the lowest RMSE, indicating the highest accuracy in force estimation. In contrast, Group 1 exhibited higher RMSE, suggesting more difficulty in accurately predicting force for taller individuals. Pearson correlation followed a similar trend, with the highest values for Group 1 and the lowest for Group 3. 

\subsection{Impact of Wearing Preferences}
Our primary study involved participants wearing both the ring and EMG watch on their dominant (right) hand, reflecting common grasping habits. To examine the effect of wearing configuration, we conducted a follow-up test where five participants wore the devices on their non-dominant (left) hand. Results remained comparable, with an average Pearson correlation coefficient (PCC) of 0.65 ± 0.08 (left) versus 0.70 ± 0.12 (right), indicating stable performance across both sides.

However, we note a practical limitation in scenarios where the EMG sensor is worn on the non-dominant hand while interactions are primarily performed with the dominant hand. In such cases, the EMG signal may not accurately reflect the active hand’s motion. Our current setup assumes sensing and interaction occur on the same limb; future work could explore cross-limb generalization or more flexible sensing configurations.

\subsection{Ablation Study}

To assess the contributions of each component in our model, we conducted an ablation study by sequentially removing key features and evaluating the impact on performance. Specifically, we explored the effects of removing the EMG channel, the IMU channel, and the cross-attention loss component. The results, summarized in Table 1, show that each of these components plays a significant role in improving the model's performance.
\paragraph{Full Model (Ours)} As a baseline, the full model achieved an RMSE of 0.213, MAE of 0.076, and a Pearson correlation of 0.759. These results demonstrate the combined effectiveness of the EMG and IMU channels along with the cross-attention loss in providing accurate finger pressure predictions.
\paragraph{w/o EMG Channel} Removing the EMG channel led to a noticeable increase in error, with RMSE rising to 0.287 and MAE increasing to 0.107. The Pearson correlation dropped to 0.628, indicating that the wrist-based EMG signal is essential for capturing fine-grained muscle activity, which significantly contributes to force estimation accuracy.
\paragraph{w/o IMU Channel} Excluding the IMU channel resulted in even worse performance, with RMSE increasing to 0.374 and MAE to 0.152. The Pearson correlation dropped sharply to 0.208, highlighting the crucial role of the IMU in capturing dynamic hand movements that are essential for accurate pressure estimation.
\paragraph{w/o Cross-Attention Loss}  when the cross-attention loss component was removed, the model's performance showed a slight improvement in RMSE (0.164) and MAE (0.132), but the Pearson correlation dropped to 0.562. This suggests that the cross-attention loss, which helps better integrate the data from different sensors, provides important contextual alignment, although its removal still leaves the model with some predictive power.

\begin{table}[!tbp]  
\centering  
\label{tab:ablation}  
\begin{tabular}{lccc}  
\hline
\textbf{Ablation Setting} & \textbf{RMSE↓} & \textbf{MAE↓} & \textbf{Pearson↑} \\
\hline
\text{Full Model (Ours)}   & 0.213 & 0.076 & 0.759 \\
\text{w/o EMG Channel}     & 0.287 & 0.107 & 0.628 \\
\text{w/o IMU Channel}         & 0.374 & 0.152 & 0.208 \\
\text{w/o Cross-Attention Loss}            & 0.164 & 0.132 & 0.562 \\
\hline
\end{tabular}
\caption{Ablation study results on different model components. "w/o" indicates removing the corresponding module.}
\end{table}


\subsection{Real-time Performance}
\paragraph{Interaction Latency} In our system, latency is critical for both VR and everyday interaction scenarios. We evaluated latency on both computer and mobile platforms to ensure the system's responsiveness across different use cases. For the desktop version, the average latency was measured at 8ms under typical usage conditions. On the mobile version (run at Vivo X90), the latency averaged at 29ms. These values were obtained by testing the system with standard finger interactions, such as pressing and releasing, as well as more complex gestures involving multiple fingers. These results demonstrate that the system meets the real-time performance requirements for VR environments, where latency should generally stay under 50ms to avoid user discomfort, and for mobile applications, where responsiveness is critical for a smooth user experience.

\paragraph{Battery Life} We evaluated the power consumption of the system to assess its suitability for prolonged use. The system is powered by two 25mAh batteries, and we conducted tests under full operational conditions. With both the IMU and EMG sensors continuously processing data, the battery life was approximately 2 hours. However, it's important to note that typical interaction durations are much shorter than the full operational time, meaning the system can last longer in practical use.

\section{Applications} 

\subsection{Visualizing Force-aware Interaction}
To demonstrate the practical applications of our force-aware interaction, we developed a Unity-based application. The user wears the Wrist2Finger ring-watch system, with real-time model predictions streamed into Unity to drive a virtual hand avatar displayed in VR. The model operates as a background process on the PC, processing live IMU and EMG data from the wearable device (sent via Bluetooth) to produce pose and force outputs at a frequency of 50 Hz. We implemented a Unity plugin that seamlessly receives and integrates these outputs.

\begin{figure}[!tbp] 
\centering \includegraphics[width=0.45\textwidth]{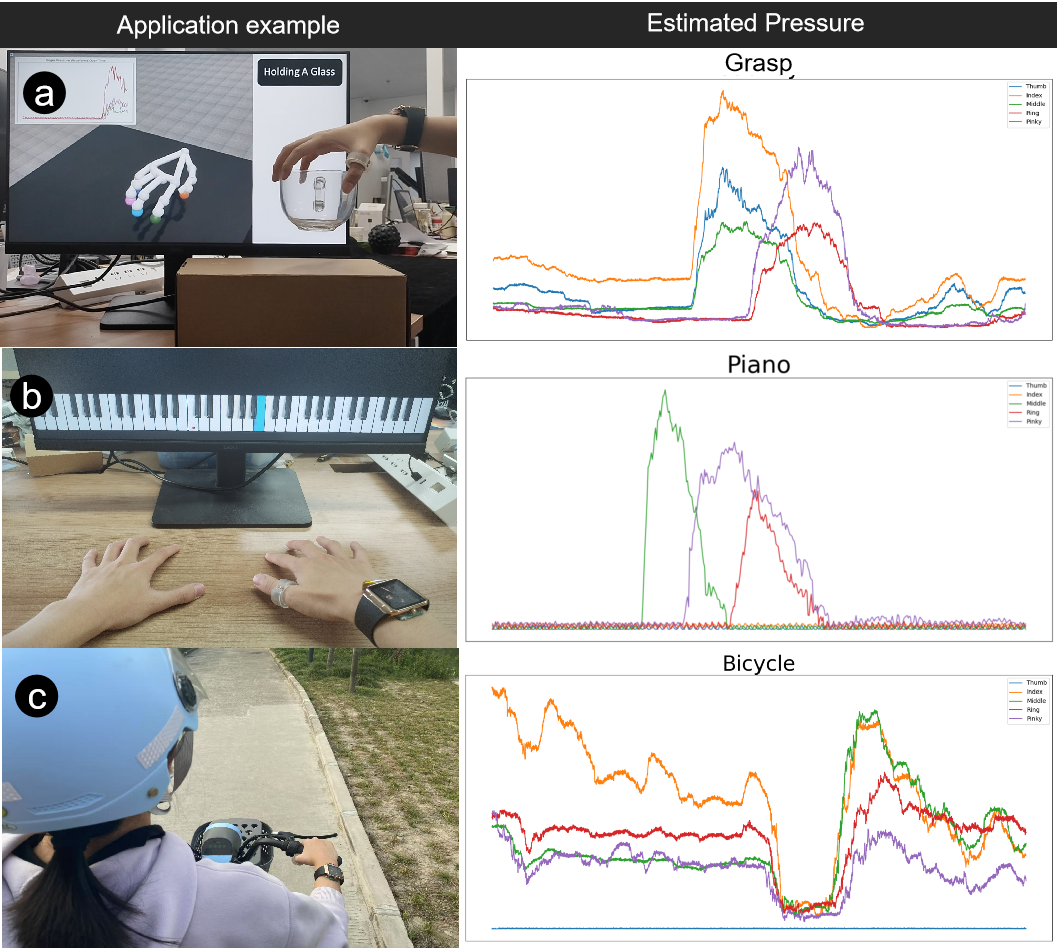}
\vspace{-10pt}
\caption{Application scenarios and the corresponding estimated pressure.} 
\label{fig:application} 
\vspace{-10pt}
\end{figure}

In Unity, as shown in Figure\ref{fig:application}(a), we employ a fully articulated hand model, consisting of 21 finger joints, which is dynamically animated based on the predicted hand pose. Additionally, force data is visualized in a separate window, providing a clear representation of the pressure applied during interaction. This setup offers an intuitive and immersive way to visualize the relationship between physical force and hand pose in real time, highlighting the system's capabilities in simulating realistic, force-aware interactions.

\subsection{Daily Interactive Scenarios}
Wrist2Finger is applicable across various daily activities where one hand must remain engaged. By leveraging wrist and finger pressure, the system facilitates uninterrupted interactions with digital devices, ensuring users remain efficient in their tasks without compromising safety or convenience. The system supports force-aware expressive interactions, enabling users to control devices through a combination of hand pose and pressure, as well as pressure-based micro-interactions that allow subtle control even when the hands are physically occupied. This force-aware interaction not only expands the usability of existing devices but also introduces a new dimension of hands-free control in everyday life.

\paragraph{Playing a Virtual Piano with Expressive Control}

Wrist2Finger enables more expressive and nuanced control in a virtual piano application, as shown in Figure~\ref{fig:application}(b). By using hand pose to select different keys and fingertip pressure to control the volume or tone, users can perform music with natural expressiveness—pressing lightly for softer notes or applying more force for louder, brighter sounds. This capability allows for an intuitive, dynamic musical experience, closely replicating real-world piano performance.

\paragraph{Cycling with Hands Occupied} When cycling, users typically maintain a firm grip on the handlebars, limiting their ability to interact with other devices. as shown in Figure\ref{fig:application}(c) Wrist2Finger allows users to control their smartphones seamlessly, such as adjusting music or volume, by applying varying pressure through the wrist. This interaction eliminates the need for users to release their grip, ensuring a safe and hands-free experience while cycling.

\paragraph{Public Transport with Handrails} In crowded environments, such as public transport, users often need to hold onto handrails for stability. Despite having one hand occupied, Wrist2Finger allows for easy manipulation of a phone or tablet by detecting pressure changes in the wrist. This functionality makes it possible to adjust music, send messages, or interact with other apps without needing to free up the hand from the handrail.

\section{Discussion and Limitation} 

Our results show that a ring-mounted IMU combined with a wrist EMG sensor can achieve hand pose and force sensing performance on par with far more complex setups. This finding is significant for the design of next-generation wearable interfaces – it suggests we don’t need full gloves or multi-electrode arrays to get rich hand input; a cleverly chosen minimal sensor set with the right AI fusion can suffice. The ring-watch approach is lightweight and compatible with daily wear, making it promising for real-world use beyond lab settings.

\subsection{Robustness and Usability}

Wrist2Finger exhibited consistently strong performance across a wide range of real-world usage scenarios, including diverse lighting conditions, varied body postures, and unconstrained hand movements. This robustness is largely attributed to its innovative camera-free design, which eliminates reliance on visual tracking and allows the system to function reliably even when the hand is partially occluded or engaged in casual, everyday gestures. For example, Wrist2Finger maintained accurate tracking even when the user’s hand was resting on a surface, momentarily placed inside a pocket, or performing loosely defined motions—situations where vision-based systems often struggle.

The system’s use of a ring-mounted IMU (Inertial Measurement Unit) provided stable and low-drift relative pose estimation over short periods, effectively capturing the dynamic motion of the fingers without significant deviation. Although the EMG (electromyography) signal quality is inherently sensitive to the quality of electrode-skin contact, the use of a snug, well-fitted wrist strap ensured consistent signal acquisition throughout typical usage. Looking ahead, future iterations could further enhance usability by adopting dry-contact electrodes or incorporating conductive textiles directly into the strap design, thereby improving comfort and reducing setup friction for everyday or long-term use.

Participants in user studies consistently described the device as lightweight, comfortable, and non-restrictive. Compared to glove-based systems, which often impede natural finger movement and introduce bulk, the single-ring form factor of Wrist2Finger provided a significantly higher degree of freedom and wearability. While the current implementation primarily focuses on gesture recognition and discrete force estimation, a promising future direction lies in the realm of continuous force control. Such functionality could unlock powerful applications, including fine-grained grip strength training for surgical skill development, or expressive pressure-based modulation in fields like digital art, music performance, and interactive gaming.

\subsection{Toward Generalization with Minimal Hardware} 

Electromyography (EMG) signals are inherently user-specific, influenced by individual muscle anatomy, skin conductivity, electrode placement, and even minor variations in posture. This variability makes cross-user generalization a central challenge for EMG-based systems. In our experiments, models trained per user achieved high accuracy, while models trained across users showed moderate but inconsistent transferability. In practice, a brief calibration session—requiring just a few standard gestures—can personalize the system, but even this step introduces friction in real-world deployment.

To reduce the need for user-specific retraining, future work could explore few-shot learning or transfer learning approaches that adapt pre-trained models using minimal new data. For example, gesture-level embeddings could be learned in a contrastive setting to encourage more general representations across users. Alternatively, meta-learning frameworks might allow rapid adaptation to new users using only one or two demonstration sequences.

A fundamental bottleneck lies in the sparse nature of single-channel EMG, which provides limited coverage of muscle activity and lacks spatial resolution. In contrast, multi-channel EMG offers richer discriminative features and better user separation, but increases hardware cost and system complexity. To balance this trade-off, we propose using knowledge distillation: a multi-channel teacher model trained offline can guide a compact single-channel student model to approximate high-resolution behavior while remaining deployable. This approach offers a practical path to enhance generalization without sacrificing form factor.

In addition, contextual information may help disambiguate ambiguous gestures or force levels. Task-aware priors—such as object affordances, interaction history, or even occasional vision input—could provide useful grounding to supplement sparse EMG data. Incorporating such high-level cues may be particularly helpful in complex or dynamic environments where raw muscle signals alone are insufficient.

\subsection{Integration with Haptics}

Accurate sensing of both hand pose and fingertip force opens up new opportunities for delivering meaningful and context-aware haptic feedback. With Wrist2Finger, the combination of IMU-based motion tracking and EMG-based force estimation provides a reliable foundation for real-time feedback generation. For instance, a ring-mounted actuator could deliver localized vibrotactile signals to simulate touch, contact, or surface texture, while the wristband could apply pressure, squeezing, or even directional cues to convey resistance or interaction force. These output channels, when coupled with precise input sensing, can significantly enhance immersion in virtual or augmented environments by simulating object stiffness, touch thresholds, or force boundaries in a natural and intuitive manner.

Crucially, Wrist2Finger enables real-time, closed-loop interaction, rather than the post hoc warnings found in many prior force-feedback systems. This closed loop allows the system to respond immediately as the user modulates grip strength or finger pressure, providing subtle corrections or confirmations during interaction. Such a capability could be leveraged in skill learning and safety-critical scenarios, including surgical simulation, fine motor training, or rehabilitation, where accurate force control is essential. Instead of passively detecting excessive force, the system could gently guide users toward optimal interaction strategies—for example, by increasing vibration intensity as grip exceeds target thresholds, or by simulating elastic resistance to encourage controlled motion.

Looking ahead, we envision future wearable systems that not only sense user behavior but actively shape it through haptic feedback. By closing the loop between sensing and actuation, Wrist2Finger could support adaptive training paradigms, in which users improve their force precision and consistency over time through continuous, embodied feedback. This opens the door to personalized motor skill refinement, expressive digital input, and safe force modulation in complex tasks. Ultimately, we see haptic integration not just as an enhancement layer, but as a core component in transforming sensing wearables into intelligent, interactive teaching tools.

\subsection{Self-Adaptation and Privacy}

While our system currently requires a short calibration phase to personalize the EMG model for each user, Wrist2Finger also opens the door to long-term, self-adaptive interaction. In everyday use, users may unconsciously adjust their gesture habits or interaction styles, especially when feedback mechanisms are involved. A static model may become suboptimal over time, leading to performance drift or degraded responsiveness. To address this, future versions could implement lightweight online learning, where the system gradually refines its internal representations based on continuous user interaction. Approaches like low-frequency model updates, gesture re-clustering, or confidence-driven fine-tuning could enable co-adaptation, allowing both the system and the user to adjust to each other over prolonged use.

Such self-personalizing capabilities are particularly important for long-term deployments in scenarios like rehabilitation, accessible computing, or productivity support. In these contexts, minimizing the need for repeated recalibration or external support improves user autonomy and system reliability. Crucially, these adaptive mechanisms must operate efficiently on-device to maintain responsiveness and power efficiency. Investigating on-device transfer learning, compressed personalization layers, or even federated model updates without raw data sharing are promising directions to ensure scalability and feasibility in the wild.

Beyond adaptability, Wrist2Finger’s sensing approach inherently respects user privacy, a critical concern in modern wearable systems. Unlike vision-based interaction methods, our system does not capture environmental imagery or spatial location data, and all sensing is limited to local physiological and motion signals. This makes it especially suitable for use in privacy-sensitive settings such as hospitals, elder care, or public AR environments. To further reinforce privacy by design, future iterations could emphasize secure, edge-only inference pipelines and explore differentially private model adaptation, ensuring that personalization never compromises user confidentiality.


\section{Conclusion}

We introduced Wrist2Finger, a minimalist ring-watch system that enables force-aware hand interactions using just a single IMU-equipped ring and a single-channel EMG sensor. Our cross-attention fusion model effectively combines kinematic and physiological signals to support both accurate hand pose reconstruction and fine-grained fingertip force estimation. Compared to vision-based and multi-device setups, Wrist2Finger offers a lightweight, comfortable, and socially acceptable alternative for real-world use. Through extensive evaluations and a VR demo, we demonstrate its potential for expressive, safe, and always-available interactions. We see Wrist2Finger as a step toward practical, scalable hand input for mobile, immersive, and assistive applications.

\begin{acks}
This work was supported by the National Natural Science Foundation of China (NSFC) under Grant 62302168.
This work was also supported in part by the National Key R\&D Program of China (No. 2022YFB4500603), Guangdong Provincial Key Laboratory of Human Digital Twin (No. 2022B1212010004); Guangzhou Basic Research Program (No. SL2023A04J00930); and Shenzhen Holdfound Foundation Endowed Professorship.
\end{acks}








\bibliographystyle{ACM-Reference-Format}
\bibliography{sample-base}


\begin{thebibliography}{59}


\ifx \showCODEN    \undefined \def \showCODEN     #1{\unskip}     \fi
\ifx \showDOI      \undefined \def \showDOI       #1{#1}\fi
\ifx \showISBNx    \undefined \def \showISBNx     #1{\unskip}     \fi
\ifx \showISBNxiii \undefined \def \showISBNxiii  #1{\unskip}     \fi
\ifx \showISSN     \undefined \def \showISSN      #1{\unskip}     \fi
\ifx \showLCCN     \undefined \def \showLCCN      #1{\unskip}     \fi
\ifx \shownote     \undefined \def \shownote      #1{#1}          \fi
\ifx \showarticletitle \undefined \def \showarticletitle #1{#1}   \fi
\ifx \showURL      \undefined \def \showURL       {\relax}        \fi
\providecommand\bibfield[2]{#2}
\providecommand\bibinfo[2]{#2}
\providecommand\natexlab[1]{#1}
\providecommand\showeprint[2][]{arXiv:#2}

\bibitem[Adafruit(2025a)]%
        {AndroidAppRef}
\bibfield{author}{\bibinfo{person}{Adafruit}.} \bibinfo{year}{2025}\natexlab{a}.
\newblock \bibinfo{booktitle}{\emph{Bluefruit LE Connect Android v2}}.
\newblock
\urldef\tempurl%
\url{https://github.com/adafruit/Bluefruit_LE_Connect_Android_v2}
\showURL{%
\tempurl}
\newblock
\shownote{Accessed: 2025-01-22}.


\bibitem[Adafruit(2025b)]%
        {IosAppRef}
\bibfield{author}{\bibinfo{person}{Adafruit}.} \bibinfo{year}{2025}\natexlab{b}.
\newblock \bibinfo{booktitle}{\emph{Bluefruit LE Connect iOS v2}}.
\newblock
\urldef\tempurl%
\url{https://github.com/adafruit/Bluefruit_LE_Connect_v2}
\showURL{%
\tempurl}
\newblock
\shownote{Accessed: 2025-01-22}.


\bibitem[Bangaru et~al\mbox{.}(2020)]%
        {bangaru2020data}
\bibfield{author}{\bibinfo{person}{Srikanth~Sagar Bangaru}, \bibinfo{person}{Chao Wang}, {and} \bibinfo{person}{Fereydoun Aghazadeh}.} \bibinfo{year}{2020}\natexlab{}.
\newblock \showarticletitle{Data quality and reliability assessment of wearable EMG and IMU sensor for construction activity recognition}.
\newblock \bibinfo{journal}{\emph{Sensors}} \bibinfo{volume}{20}, \bibinfo{number}{18} (\bibinfo{year}{2020}), \bibinfo{pages}{5264}.
\newblock


\bibitem[Bhiri et~al\mbox{.}(2023)]%
        {MajdoubBhiri2023HandGR}
\bibfield{author}{\bibinfo{person}{Nahla~Majdoub Bhiri}, \bibinfo{person}{Safa Ameur}, \bibinfo{person}{Ihsen Alouani}, \bibinfo{person}{Mohamed~Ali Mahjoub}, {and} \bibinfo{person}{Anouar~Ben Khalifa}.} \bibinfo{year}{2023}\natexlab{}.
\newblock \showarticletitle{Hand gesture recognition with focus on leap motion: An overview, real world challenges and future directions}.
\newblock \bibinfo{journal}{\emph{Expert Syst. Appl.}}  \bibinfo{volume}{226} (\bibinfo{year}{2023}), \bibinfo{pages}{120125}.
\newblock
\urldef\tempurl%
\url{https://api.semanticscholar.org/CorpusID:258224705}
\showURL{%
\tempurl}


\bibitem[Cao et~al\mbox{.}(2024)]%
        {10623717}
\bibfield{author}{\bibinfo{person}{Jiani Cao}, \bibinfo{person}{Yang Liu}, \bibinfo{person}{Lixiang Han}, {and} \bibinfo{person}{Zhenjiang Li}.} \bibinfo{year}{2024}\natexlab{}.
\newblock \showarticletitle{Finger Tracking Using Wrist-Worn EMG Sensors}.
\newblock \bibinfo{journal}{\emph{IEEE Transactions on Mobile Computing}} \bibinfo{volume}{23}, \bibinfo{number}{12} (\bibinfo{year}{2024}), \bibinfo{pages}{14099--14110}.
\newblock
\urldef\tempurl%
\url{https://doi.org/10.1109/TMC.2024.3439018}
\showDOI{\tempurl}


\bibitem[Cho and Kim(2022)]%
        {cho2022real}
\bibfield{author}{\bibinfo{person}{Younggeol Cho} {and} \bibinfo{person}{Pyungkang Kim}.} \bibinfo{year}{2022}\natexlab{}.
\newblock \showarticletitle{Real-time finger force estimation robust to a perturbation of electrode placement for prosthetic hand control}.
\newblock \bibinfo{journal}{\emph{IEEE Transactions on Neural Systems and Rehabilitation Engineering}}  \bibinfo{volume}{30} (\bibinfo{year}{2022}), \bibinfo{pages}{1233--1243}.
\newblock


\bibitem[Connan et~al\mbox{.}(2016)]%
        {connan2016assessment}
\bibfield{author}{\bibinfo{person}{Mathilde Connan}, \bibinfo{person}{Eduardo Ruiz~Ram{\'\i}rez}, \bibinfo{person}{Bernhard Vodermayer}, {and} \bibinfo{person}{Claudio Castellini}.} \bibinfo{year}{2016}\natexlab{}.
\newblock \showarticletitle{Assessment of a wearable force-and electromyography device and comparison of the related signals for myocontrol}.
\newblock \bibinfo{journal}{\emph{Frontiers in neurorobotics}}  \bibinfo{volume}{10} (\bibinfo{year}{2016}), \bibinfo{pages}{17}.
\newblock


\bibitem[Dong et~al\mbox{.}(2021)]%
        {Dong2021DynamicHG}
\bibfield{author}{\bibinfo{person}{Yongfeng Dong}, \bibinfo{person}{Jie Liu}, {and} \bibinfo{person}{Wenjie Yan}.} \bibinfo{year}{2021}\natexlab{}.
\newblock \showarticletitle{Dynamic Hand Gesture Recognition Based on Signals From Specialized Data Glove and Deep Learning Algorithms}.
\newblock \bibinfo{journal}{\emph{IEEE Transactions on Instrumentation and Measurement}}  \bibinfo{volume}{70} (\bibinfo{year}{2021}), \bibinfo{pages}{1--14}.
\newblock
\urldef\tempurl%
\url{https://api.semanticscholar.org/CorpusID:234787591}
\showURL{%
\tempurl}


\bibitem[Du et~al\mbox{.}(2022)]%
        {Du2022AMG}
\bibfield{author}{\bibinfo{person}{Guanglong Du}, \bibinfo{person}{Dawei Guo}, \bibinfo{person}{Kang Su}, \bibinfo{person}{Xueqian Wang}, \bibinfo{person}{Shaohua Teng}, \bibinfo{person}{Di Li}, {and} \bibinfo{person}{Peter~Xiaoping Liu}.} \bibinfo{year}{2022}\natexlab{}.
\newblock \showarticletitle{A Mobile Gesture Interaction Method for Augmented Reality Games Using Hybrid Filters}.
\newblock \bibinfo{journal}{\emph{IEEE Transactions on Instrumentation and Measurement}}  \bibinfo{volume}{71} (\bibinfo{year}{2022}), \bibinfo{pages}{1--12}.
\newblock
\urldef\tempurl%
\url{https://api.semanticscholar.org/CorpusID:249664979}
\showURL{%
\tempurl}


\bibitem[Gao et~al\mbox{.}(2023)]%
        {9903078}
\bibfield{author}{\bibinfo{person}{Qing Gao}, \bibinfo{person}{Zhaojie Ju}, \bibinfo{person}{Yongquan Chen}, \bibinfo{person}{Qiwen Wang}, {and} \bibinfo{person}{Chuliang Chi}.} \bibinfo{year}{2023}\natexlab{}.
\newblock \showarticletitle{An Efficient RGB-D Hand Gesture Detection Framework for Dexterous Robot Hand-Arm Teleoperation System}.
\newblock \bibinfo{journal}{\emph{IEEE Transactions on Human-Machine Systems}} \bibinfo{volume}{53}, \bibinfo{number}{1} (\bibinfo{year}{2023}), \bibinfo{pages}{13--23}.
\newblock
\urldef\tempurl%
\url{https://doi.org/10.1109/THMS.2022.3206663}
\showDOI{\tempurl}


\bibitem[Georgi et~al\mbox{.}(2015)]%
        {georgi2015fusion}
\bibfield{author}{\bibinfo{person}{Marcus Georgi}, \bibinfo{person}{Christoph Amma}, {and} \bibinfo{person}{Tanja Schultz}.} \bibinfo{year}{2015}\natexlab{}.
\newblock \showarticletitle{Fusion and Comparison of IMU and EMG signals for wearable gesture recognition}. In \bibinfo{booktitle}{\emph{International Joint Conference on Biomedical Engineering Systems and Technologies}}. Springer, \bibinfo{pages}{308--323}.
\newblock


\bibitem[Google(2025)]%
        {mediapipe}
\bibfield{author}{\bibinfo{person}{Google}.} \bibinfo{year}{2025}\natexlab{}.
\newblock \bibinfo{title}{Hand landmarks detection guide}.
\newblock
\newblock
\urldef\tempurl%
\url{https://ai.google.dev/edge/mediapipe/solutions/vision/hand_landmarker}
\showURL{%
Retrieved April 7, 2025 from \tempurl}


\bibitem[Gosala et~al\mbox{.}(2023)]%
        {9628050}
\bibfield{author}{\bibinfo{person}{Nikhil Gosala}, \bibinfo{person}{Fangjinhua Wang}, \bibinfo{person}{Zhaopeng Cui}, \bibinfo{person}{Hanxue Liang}, \bibinfo{person}{Oliver Glauser}, \bibinfo{person}{Shihao Wu}, {and} \bibinfo{person}{Olga Sorkine-Hornung}.} \bibinfo{year}{2023}\natexlab{}.
\newblock \showarticletitle{Self-Calibrated Multi-Sensor Wearable for Hand Tracking and Modeling}.
\newblock \bibinfo{journal}{\emph{IEEE Transactions on Visualization and Computer Graphics}} \bibinfo{volume}{29}, \bibinfo{number}{3} (\bibinfo{year}{2023}), \bibinfo{pages}{1769--1784}.
\newblock
\urldef\tempurl%
\url{https://doi.org/10.1109/TVCG.2021.3131230}
\showDOI{\tempurl}


\bibitem[Huo et~al\mbox{.}(2021)]%
        {leapmotion}
\bibfield{author}{\bibinfo{person}{Jiage Huo}, \bibinfo{person}{K.~L. Keung}, \bibinfo{person}{C. Lee}, {and} \bibinfo{person}{H. Ng}.} \bibinfo{year}{2021}\natexlab{}.
\newblock \showarticletitle{Hand Gesture Recognition with Augmented Reality and Leap Motion Controller}. \bibinfo{pages}{1015--1019}.
\newblock
\urldef\tempurl%
\url{https://doi.org/10.1109/IEEM50564.2021.9672611}
\showDOI{\tempurl}


\bibitem[Jiang et~al\mbox{.}(2017)]%
        {jiang2017feasibility}
\bibfield{author}{\bibinfo{person}{Shuo Jiang}, \bibinfo{person}{Bo Lv}, \bibinfo{person}{Weichao Guo}, \bibinfo{person}{Chao Zhang}, \bibinfo{person}{Haitao Wang}, \bibinfo{person}{Xinjun Sheng}, {and} \bibinfo{person}{Peter~B Shull}.} \bibinfo{year}{2017}\natexlab{}.
\newblock \showarticletitle{Feasibility of wrist-worn, real-time hand, and surface gesture recognition via sEMG and IMU sensing}.
\newblock \bibinfo{journal}{\emph{IEEE Transactions on Industrial Informatics}} \bibinfo{volume}{14}, \bibinfo{number}{8} (\bibinfo{year}{2017}), \bibinfo{pages}{3376--3385}.
\newblock


\bibitem[Karheily et~al\mbox{.}(2022)]%
        {Karheily2022sEMGTF}
\bibfield{author}{\bibinfo{person}{Somar Karheily}, \bibinfo{person}{Ali Moukadem}, \bibinfo{person}{Jean-Baptiste Courbot}, {and} \bibinfo{person}{Djaffar~Ould Abdeslam}.} \bibinfo{year}{2022}\natexlab{}.
\newblock \showarticletitle{sEMG time-frequency features for hand movements classification}.
\newblock \bibinfo{journal}{\emph{Expert Syst. Appl.}}  \bibinfo{volume}{210} (\bibinfo{year}{2022}), \bibinfo{pages}{118282}.
\newblock
\urldef\tempurl%
\url{https://api.semanticscholar.org/CorpusID:251203863}
\showURL{%
\tempurl}


\bibitem[Khomami and Shamekhi(2021)]%
        {Khomami2021PersianSL}
\bibfield{author}{\bibinfo{person}{Sara~Askari Khomami} {and} \bibinfo{person}{Sina Shamekhi}.} \bibinfo{year}{2021}\natexlab{}.
\newblock \showarticletitle{Persian sign language recognition using IMU and surface EMG sensors}.
\newblock \bibinfo{journal}{\emph{Measurement}}  \bibinfo{volume}{168} (\bibinfo{year}{2021}), \bibinfo{pages}{108471}.
\newblock
\urldef\tempurl%
\url{https://api.semanticscholar.org/CorpusID:225024512}
\showURL{%
\tempurl}


\bibitem[Kim and Harrison(2022)]%
        {10.1145/3526113.3545665}
\bibfield{author}{\bibinfo{person}{Daehwa Kim} {and} \bibinfo{person}{Chris Harrison}.} \bibinfo{year}{2022}\natexlab{}.
\newblock \showarticletitle{EtherPose: Continuous Hand Pose Tracking with Wrist-Worn Antenna Impedance Characteristic Sensing}. In \bibinfo{booktitle}{\emph{Proceedings of the 35th Annual ACM Symposium on User Interface Software and Technology}} (Bend, OR, USA) \emph{(\bibinfo{series}{UIST '22})}. \bibinfo{publisher}{Association for Computing Machinery}, \bibinfo{address}{New York, NY, USA}, Article \bibinfo{articleno}{58}, \bibinfo{numpages}{12}~pages.
\newblock
\showISBNx{9781450393201}


\bibitem[Kyu et~al\mbox{.}(2024)]%
        {10.1145/3613904.3642663}
\bibfield{author}{\bibinfo{person}{Alexander Kyu}, \bibinfo{person}{Hongyu Mao}, \bibinfo{person}{Junyi Zhu}, \bibinfo{person}{Mayank Goel}, {and} \bibinfo{person}{Karan Ahuja}.} \bibinfo{year}{2024}\natexlab{}.
\newblock \showarticletitle{EITPose: Wearable and Practical Electrical Impedance Tomography for Continuous Hand Pose Estimation}. In \bibinfo{booktitle}{\emph{Proceedings of the 2024 CHI Conference on Human Factors in Computing Systems}} (Honolulu, HI, USA) \emph{(\bibinfo{series}{CHI '24})}. \bibinfo{publisher}{Association for Computing Machinery}, \bibinfo{address}{New York, NY, USA}, Article \bibinfo{articleno}{402}, \bibinfo{numpages}{10}~pages.
\newblock
\showISBNx{9798400703300}


\bibitem[Lee et~al\mbox{.}(2024)]%
        {10.1145/3613904.3642910}
\bibfield{author}{\bibinfo{person}{Chi-Jung Lee}, \bibinfo{person}{Ruidong Zhang}, \bibinfo{person}{Devansh Agarwal}, \bibinfo{person}{Tianhong~Catherine Yu}, \bibinfo{person}{Vipin Gunda}, \bibinfo{person}{Oliver Lopez}, \bibinfo{person}{James Kim}, \bibinfo{person}{Sicheng Yin}, \bibinfo{person}{Boao Dong}, \bibinfo{person}{Ke Li}, \bibinfo{person}{Mose Sakashita}, \bibinfo{person}{Francois Guimbretiere}, {and} \bibinfo{person}{Cheng Zhang}.} \bibinfo{year}{2024}\natexlab{}.
\newblock \showarticletitle{EchoWrist: Continuous Hand Pose Tracking and Hand-Object Interaction Recognition Using Low-Power Active Acoustic Sensing On a Wristband}. In \bibinfo{booktitle}{\emph{Proceedings of the 2024 CHI Conference on Human Factors in Computing Systems}} (Honolulu, HI, USA) \emph{(\bibinfo{series}{CHI '24})}. \bibinfo{publisher}{Association for Computing Machinery}, \bibinfo{address}{New York, NY, USA}, Article \bibinfo{articleno}{403}, \bibinfo{numpages}{21}~pages.
\newblock
\showISBNx{9798400703300}


\bibitem[Li et~al\mbox{.}(2013)]%
        {li2013directional}
\bibfield{author}{\bibinfo{person}{Ke Li}, \bibinfo{person}{Raviraj Nataraj}, \bibinfo{person}{Tamara~L Marquardt}, {and} \bibinfo{person}{Zong-Ming Li}.} \bibinfo{year}{2013}\natexlab{}.
\newblock \showarticletitle{Directional coordination of thumb and finger forces during precision pinch}.
\newblock \bibinfo{journal}{\emph{PloS one}} \bibinfo{volume}{8}, \bibinfo{number}{11} (\bibinfo{year}{2013}), \bibinfo{pages}{e79400}.
\newblock


\bibitem[Liang et~al\mbox{.}(2021a)]%
        {liang2021dualring}
\bibfield{author}{\bibinfo{person}{Chen Liang}, \bibinfo{person}{Chun Yu}, \bibinfo{person}{Yue Qin}, \bibinfo{person}{Yuntao Wang}, {and} \bibinfo{person}{Yuanchun Shi}.} \bibinfo{year}{2021}\natexlab{a}.
\newblock \showarticletitle{DualRing: Enabling subtle and expressive hand interaction with dual IMU rings}.
\newblock \bibinfo{journal}{\emph{Proceedings of the ACM on Interactive, Mobile, Wearable and Ubiquitous Technologies}} \bibinfo{volume}{5}, \bibinfo{number}{3} (\bibinfo{year}{2021}), \bibinfo{pages}{1--27}.
\newblock


\bibitem[Liang et~al\mbox{.}(2021b)]%
        {DBLP}
\bibfield{author}{\bibinfo{person}{Chen Liang}, \bibinfo{person}{Chun Yu}, \bibinfo{person}{Yue Qin}, \bibinfo{person}{Yuntao Wang}, {and} \bibinfo{person}{Yuanchun Shi}.} \bibinfo{year}{2021}\natexlab{b}.
\newblock \showarticletitle{DualRing: Enabling Subtle and Expressive Hand Interaction with Dual {IMU} Rings}.
\newblock \bibinfo{journal}{\emph{Proc. {ACM} Interact. Mob. Wearable Ubiquitous Technol.}} \bibinfo{volume}{5}, \bibinfo{number}{3} (\bibinfo{year}{2021}), \bibinfo{pages}{115:1--115:27}.
\newblock
\urldef\tempurl%
\url{https://doi.org/10.1145/3478114}
\showDOI{\tempurl}


\bibitem[Lieber et~al\mbox{.}(1992)]%
        {lieber1992architecture}
\bibfield{author}{\bibinfo{person}{Richard~L Lieber}, \bibinfo{person}{Mark~D Jacobson}, \bibinfo{person}{Babak~M Fazeli}, \bibinfo{person}{Reid~A Abrams}, {and} \bibinfo{person}{Michael~J Botte}.} \bibinfo{year}{1992}\natexlab{}.
\newblock \showarticletitle{Architecture of selected muscles of the arm and forearm: anatomy and implications for tendon transfer}.
\newblock \bibinfo{journal}{\emph{The Journal of hand surgery}} \bibinfo{volume}{17}, \bibinfo{number}{5} (\bibinfo{year}{1992}), \bibinfo{pages}{787--798}.
\newblock


\bibitem[Liu et~al\mbox{.}(2021a)]%
        {article}
\bibfield{author}{\bibinfo{person}{Yang Liu}, \bibinfo{person}{Chengdong Lin}, {and} \bibinfo{person}{Zhenjiang Li}.} \bibinfo{year}{2021}\natexlab{a}.
\newblock \showarticletitle{WR-Hand: Wearable Armband Can Track User's Hand}.
\newblock \bibinfo{journal}{\emph{Proceedings of the ACM on Interactive, Mobile, Wearable and Ubiquitous Technologies}}  \bibinfo{volume}{5} (\bibinfo{date}{09} \bibinfo{year}{2021}), \bibinfo{pages}{1--27}.
\newblock
\urldef\tempurl%
\url{https://doi.org/10.1145/3478112}
\showDOI{\tempurl}


\bibitem[Liu et~al\mbox{.}(2021b)]%
        {liu2021neuropose}
\bibfield{author}{\bibinfo{person}{Yilin Liu}, \bibinfo{person}{Shijia Zhang}, {and} \bibinfo{person}{Mahanth Gowda}.} \bibinfo{year}{2021}\natexlab{b}.
\newblock \showarticletitle{NeuroPose: 3D hand pose tracking using EMG wearables}. In \bibinfo{booktitle}{\emph{Proceedings of the Web Conference 2021}}. \bibinfo{pages}{1471--1482}.
\newblock


\bibitem[Liu et~al\mbox{.}(2023)]%
        {9956792}
\bibfield{author}{\bibinfo{person}{Yilin Liu}, \bibinfo{person}{Shijia Zhang}, {and} \bibinfo{person}{Mahanth Gowda}.} \bibinfo{year}{2023}\natexlab{}.
\newblock \showarticletitle{A Practical System for 3-D Hand Pose Tracking Using EMG Wearables With Applications to Prosthetics and User Interfaces}.
\newblock \bibinfo{journal}{\emph{IEEE Internet of Things Journal}} \bibinfo{volume}{10}, \bibinfo{number}{4} (\bibinfo{year}{2023}), \bibinfo{pages}{3407--3427}.
\newblock
\urldef\tempurl%
\url{https://doi.org/10.1109/JIOT.2022.3223600}
\showDOI{\tempurl}


\bibitem[Lopes et~al\mbox{.}(2017)]%
        {lopes2017hand}
\bibfield{author}{\bibinfo{person}{Jo{\~a}o Lopes}, \bibinfo{person}{Miguel Sim{\~a}o}, \bibinfo{person}{Nuno Mendes}, \bibinfo{person}{Mohammad Safeea}, \bibinfo{person}{Jos{\'e} Afonso}, {and} \bibinfo{person}{Pedro Neto}.} \bibinfo{year}{2017}\natexlab{}.
\newblock \showarticletitle{Hand/arm gesture segmentation by motion using IMU and EMG sensing}.
\newblock \bibinfo{journal}{\emph{Procedia Manufacturing}}  \bibinfo{volume}{11} (\bibinfo{year}{2017}), \bibinfo{pages}{107--113}.
\newblock


\bibitem[Mao et~al\mbox{.}(2023)]%
        {mao2023simultaneous}
\bibfield{author}{\bibinfo{person}{He Mao}, \bibinfo{person}{Yue Zheng}, \bibinfo{person}{Chenfei Ma}, \bibinfo{person}{Kai Wu}, \bibinfo{person}{Guanglin Li}, {and} \bibinfo{person}{Peng Fang}.} \bibinfo{year}{2023}\natexlab{}.
\newblock \showarticletitle{Simultaneous estimation of grip force and wrist angles by surface electromyography and acceleration signals}.
\newblock \bibinfo{journal}{\emph{Biomedical Signal Processing and Control}}  \bibinfo{volume}{79} (\bibinfo{year}{2023}), \bibinfo{pages}{104088}.
\newblock


\bibitem[Meng and Hu(2024)]%
        {meng2024unsupervised}
\bibfield{author}{\bibinfo{person}{Long Meng} {and} \bibinfo{person}{Xiaogang Hu}.} \bibinfo{year}{2024}\natexlab{}.
\newblock \showarticletitle{Unsupervised neural decoding for concurrent and continuous multi-finger force prediction}.
\newblock \bibinfo{journal}{\emph{Computers in Biology and Medicine}}  \bibinfo{volume}{173} (\bibinfo{year}{2024}), \bibinfo{pages}{108384}.
\newblock


\bibitem[Mueller et~al\mbox{.}(2019)]%
        {mueller2019real}
\bibfield{author}{\bibinfo{person}{Franziska Mueller}, \bibinfo{person}{Micah Davis}, \bibinfo{person}{Florian Bernard}, \bibinfo{person}{Oleksandr Sotnychenko}, \bibinfo{person}{Mickeal Verschoor}, \bibinfo{person}{Miguel~A Otaduy}, \bibinfo{person}{Dan Casas}, {and} \bibinfo{person}{Christian Theobalt}.} \bibinfo{year}{2019}\natexlab{}.
\newblock \showarticletitle{Real-time pose and shape reconstruction of two interacting hands with a single depth camera}.
\newblock \bibinfo{journal}{\emph{ACM Transactions on Graphics (ToG)}} \bibinfo{volume}{38}, \bibinfo{number}{4} (\bibinfo{year}{2019}), \bibinfo{pages}{1--13}.
\newblock


\bibitem[Mueller et~al\mbox{.}(2017)]%
        {mueller2017real}
\bibfield{author}{\bibinfo{person}{Franziska Mueller}, \bibinfo{person}{Dushyant Mehta}, \bibinfo{person}{Oleksandr Sotnychenko}, \bibinfo{person}{Srinath Sridhar}, \bibinfo{person}{Dan Casas}, {and} \bibinfo{person}{Christian Theobalt}.} \bibinfo{year}{2017}\natexlab{}.
\newblock \showarticletitle{Real-time hand tracking under occlusion from an egocentric rgb-d sensor}. In \bibinfo{booktitle}{\emph{Proceedings of the IEEE international conference on computer vision}}. \bibinfo{pages}{1154--1163}.
\newblock


\bibitem[Olsson et~al\mbox{.}(2021)]%
        {olsson2021end}
\bibfield{author}{\bibinfo{person}{Alexander~E Olsson}, \bibinfo{person}{Neboj{\v{s}}a Male{\v{s}}evi{\'c}}, \bibinfo{person}{Anders Bj{\"o}rkman}, {and} \bibinfo{person}{Christian Antfolk}.} \bibinfo{year}{2021}\natexlab{}.
\newblock \showarticletitle{End-to-end estimation of hand-and wrist forces from raw intramuscular emg signals using lstm networks}.
\newblock \bibinfo{journal}{\emph{Frontiers in Neuroscience}}  \bibinfo{volume}{15} (\bibinfo{year}{2021}), \bibinfo{pages}{777329}.
\newblock


\bibitem[Pan et~al\mbox{.}(2021)]%
        {2021Hybrid}
\bibfield{author}{\bibinfo{person}{J. Pan}, \bibinfo{person}{Y. Li}, \bibinfo{person}{Y. Luo}, \bibinfo{person}{X. Zhang}, \bibinfo{person}{X. Wang}, \bibinfo{person}{Dlt. Wong}, \bibinfo{person}{Ch. Heng}, \bibinfo{person}{Ck. Tham}, {and} \bibinfo{person}{Av. Thean}.} \bibinfo{year}{2021}\natexlab{}.
\newblock \showarticletitle{Hybrid-Flexible Bimodal Sensing Wearable Glove System for Complex Hand Gesture Recognition.}
\newblock \bibinfo{journal}{\emph{ACS sensors}} \bibinfo{volume}{6}, \bibinfo{number}{11} (\bibinfo{year}{2021}), \bibinfo{pages}{4156--4166}.
\newblock


\bibitem[Parate et~al\mbox{.}(2014)]%
        {Parate2014RisQRS}
\bibfield{author}{\bibinfo{person}{Abhinav Parate}, \bibinfo{person}{Meng-Chieh Chiu}, \bibinfo{person}{Chaniel Chadowitz}, \bibinfo{person}{Deepak Ganesan}, {and} \bibinfo{person}{Evangelos Kalogerakis}.} \bibinfo{year}{2014}\natexlab{}.
\newblock \showarticletitle{RisQ: recognizing smoking gestures with inertial sensors on a wristband}.
\newblock \bibinfo{journal}{\emph{Proceedings of the 12th annual international conference on Mobile systems, applications, and services}} (\bibinfo{year}{2014}).
\newblock
\urldef\tempurl%
\url{https://api.semanticscholar.org/CorpusID:901628}
\showURL{%
\tempurl}


\bibitem[Semiconductor(2022)]%
        {nrf52832Ref}
\bibfield{author}{\bibinfo{person}{Nordic Semiconductor}.} \bibinfo{year}{2022}\natexlab{}.
\newblock \bibinfo{booktitle}{\emph{nRF52832}}.
\newblock
\urldef\tempurl%
\url{https://www.nordicsemi.com/products/nrf52832}
\showURL{%
\tempurl}
\newblock
\shownote{Accessed: 2025-01-22}.


\bibitem[Semiconductor(2025)]%
        {NordicAppRef}
\bibfield{author}{\bibinfo{person}{Nordic Semiconductor}.} \bibinfo{year}{2025}\natexlab{}.
\newblock \bibinfo{booktitle}{\emph{Android DFU Library}}.
\newblock
\urldef\tempurl%
\url{https://github.com/NordicSemiconductor/Android-DFU-Library}
\showURL{%
\tempurl}
\newblock
\shownote{Accessed: 2025-01-22}.


\bibitem[Seo et~al\mbox{.}(2024)]%
        {seo2024postureinformed}
\bibfield{author}{\bibinfo{person}{Kyungjin Seo}, \bibinfo{person}{Junghoon Seo}, \bibinfo{person}{Hanseok Jeong}, \bibinfo{person}{Sangpil Kim}, {and} \bibinfo{person}{Sang~Ho Yoon}.} \bibinfo{year}{2024}\natexlab{}.
\newblock \showarticletitle{Posture-Informed Muscular Force Learning for Robust Hand Pressure Estimation}. In \bibinfo{booktitle}{\emph{The Thirty-eighth Annual Conference on Neural Information Processing Systems}}.
\newblock


\bibitem[Siddiqui and Chan(2020)]%
        {siddiqui2020multimodal}
\bibfield{author}{\bibinfo{person}{Nabeel Siddiqui} {and} \bibinfo{person}{Rosa~HM Chan}.} \bibinfo{year}{2020}\natexlab{}.
\newblock \showarticletitle{Multimodal hand gesture recognition using single IMU and acoustic measurements at wrist}.
\newblock \bibinfo{journal}{\emph{Plos one}} \bibinfo{volume}{15}, \bibinfo{number}{1} (\bibinfo{year}{2020}), \bibinfo{pages}{e0227039}.
\newblock


\bibitem[Su et~al\mbox{.}(2021)]%
        {su2021deep}
\bibfield{author}{\bibinfo{person}{Hang Su}, \bibinfo{person}{Wen Qi}, \bibinfo{person}{Zhijun Li}, \bibinfo{person}{Ziyang Chen}, \bibinfo{person}{Giancarlo Ferrigno}, {and} \bibinfo{person}{Elena De~Momi}.} \bibinfo{year}{2021}\natexlab{}.
\newblock \showarticletitle{Deep neural network approach in EMG-based force estimation for human--robot interaction}.
\newblock \bibinfo{journal}{\emph{IEEE Transactions on Artificial Intelligence}} \bibinfo{volume}{2}, \bibinfo{number}{5} (\bibinfo{year}{2021}), \bibinfo{pages}{404--412}.
\newblock


\bibitem[Takahashi et~al\mbox{.}(2024)]%
        {takahashi2024picoring}
\bibfield{author}{\bibinfo{person}{Ryo Takahashi}, \bibinfo{person}{Eric Whitmire}, \bibinfo{person}{Roger Boldu}, \bibinfo{person}{Shiu Ng}, \bibinfo{person}{Wolf Kienzle}, {and} \bibinfo{person}{Hrvoje Benko}.} \bibinfo{year}{2024}\natexlab{}.
\newblock \showarticletitle{picoRing: battery-free rings for subtle thumb-to-index input}. In \bibinfo{booktitle}{\emph{Proceedings of the 37th Annual ACM Symposium on User Interface Software and Technology}}. \bibinfo{pages}{1--11}.
\newblock


\bibitem[Tanweer and Halonen(2019)]%
        {tanweer2019development}
\bibfield{author}{\bibinfo{person}{Muhammad Tanweer} {and} \bibinfo{person}{Kari~AI Halonen}.} \bibinfo{year}{2019}\natexlab{}.
\newblock \showarticletitle{Development of wearable hardware platform to measure the ECG and EMG with IMU to detect motion artifacts}. In \bibinfo{booktitle}{\emph{2019 IEEE 22nd International Symposium on Design and Diagnostics of Electronic Circuits \& Systems (DDECS)}}. IEEE, \bibinfo{pages}{1--4}.
\newblock


\bibitem[(TDK)(2025)]%
        {ICM20948Ref}
\bibfield{author}{\bibinfo{person}{InvenSense (TDK)}.} \bibinfo{year}{2025}\natexlab{}.
\newblock \bibinfo{booktitle}{\emph{ICM-20948: 9-Axis MotionTracking Device}}.
\newblock
\urldef\tempurl%
\url{https://invensense.tdk.com/products/motion-tracking/9-axis/icm-20948/}
\showURL{%
\tempurl}
\newblock
\shownote{Accessed: 2025-01-22}.


\bibitem[tekscan(2024)]%
        {flexiforce}
\bibfield{author}{\bibinfo{person}{tekscan}.} \bibinfo{year}{2024}\natexlab{}.
\newblock \bibinfo{title}{FlexiForce Load/Force Sensors and Systems}.
\newblock
\newblock
\urldef\tempurl%
\url{https://www.tekscan.com/flexiforce-loadforce-sensors-and-systems}
\showURL{%
Retrieved April 7, 2025 from \tempurl}


\bibitem[V{\'a}sconez et~al\mbox{.}(2022)]%
        {vasconez2022hand}
\bibfield{author}{\bibinfo{person}{Juan~Pablo V{\'a}sconez}, \bibinfo{person}{Lorena~Isabel Barona~L{\'o}pez}, \bibinfo{person}{{\'A}ngel~Leonardo Valdivieso~Caraguay}, {and} \bibinfo{person}{Marco~E Benalc{\'a}zar}.} \bibinfo{year}{2022}\natexlab{}.
\newblock \showarticletitle{Hand gesture recognition using EMG-IMU signals and deep q-networks}.
\newblock \bibinfo{journal}{\emph{Sensors}} \bibinfo{volume}{22}, \bibinfo{number}{24} (\bibinfo{year}{2022}), \bibinfo{pages}{9613}.
\newblock


\bibitem[Waghmare et~al\mbox{.}(2023)]%
        {inproceedings}
\bibfield{author}{\bibinfo{person}{Anandghan Waghmare}, \bibinfo{person}{Ishan Chatterjee}, {and} \bibinfo{person}{Shwetak Patel}.} \bibinfo{year}{2023}\natexlab{}.
\newblock \showarticletitle{Z-Pose: Continuous 3D Hand Pose Tracking Using Single-Point Bio-Impedance Sensing on a Ring}. \bibinfo{pages}{1--6}.
\newblock
\urldef\tempurl%
\url{https://doi.org/10.1145/3615592.3616851}
\showDOI{\tempurl}


\bibitem[Wahid et~al\mbox{.}(2024)]%
        {wahid2024semg}
\bibfield{author}{\bibinfo{person}{Abdul Wahid}, \bibinfo{person}{Khalil Ullah}, \bibinfo{person}{Syed~Irfan Ullah}, \bibinfo{person}{Muhammad Amin}, \bibinfo{person}{Sulaiman Almutairi}, {and} \bibinfo{person}{Mohammed Abohashrh}.} \bibinfo{year}{2024}\natexlab{}.
\newblock \showarticletitle{sEMG-Based Upper Limb Elbow Force Estimation Using CNN, CNN-LSTM, and CNN-GRU Models}.
\newblock \bibinfo{journal}{\emph{IEEE Access}} (\bibinfo{year}{2024}).
\newblock


\bibitem[Wen et~al\mbox{.}(2021)]%
        {signLanguage}
\bibfield{author}{\bibinfo{person}{Feng Wen}, \bibinfo{person}{Zixuan Zhang}, \bibinfo{person}{Tianyiyi He}, {and} \bibinfo{person}{Chengkuo Lee}.} \bibinfo{year}{2021}\natexlab{}.
\newblock \showarticletitle{AI enabled sign language recognition and VR space bidirectional communication using triboelectric smart glove}.
\newblock \bibinfo{journal}{\emph{Nature Communications}}  \bibinfo{volume}{12} (\bibinfo{date}{09} \bibinfo{year}{2021}), \bibinfo{pages}{5378}.
\newblock
\urldef\tempurl%
\url{https://doi.org/10.1038/s41467-021-25637-w}
\showDOI{\tempurl}


\bibitem[Wilhelm et~al\mbox{.}(2020)]%
        {wilhelm2020ring}
\bibfield{author}{\bibinfo{person}{Mathias Wilhelm}, \bibinfo{person}{Jan-Peter Lechler}, \bibinfo{person}{Daniel Krakowczyk}, {and} \bibinfo{person}{Sahin Albayrak}.} \bibinfo{year}{2020}\natexlab{}.
\newblock \showarticletitle{Ring-based finger tracking using capacitive sensors and long short-term memory}. In \bibinfo{booktitle}{\emph{Proceedings of the 25th International Conference on Intelligent User Interfaces}}. \bibinfo{pages}{551--555}.
\newblock


\bibitem[Wu et~al\mbox{.}(2022)]%
        {Wu2022FullFiberAY}
\bibfield{author}{\bibinfo{person}{Ronghui Wu}, \bibinfo{person}{Sangjin Seo}, \bibinfo{person}{Liyun Ma}, \bibinfo{person}{Juyeol Bae}, {and} \bibinfo{person}{Taesung Kim}.} \bibinfo{year}{2022}\natexlab{}.
\newblock \showarticletitle{Full-Fiber Auxetic-Interlaced Yarn Sensor for Sign-Language Translation Glove Assisted by Artificial Neural Network}.
\newblock \bibinfo{journal}{\emph{Nano-Micro Letters}}  \bibinfo{volume}{14} (\bibinfo{year}{2022}).
\newblock
\urldef\tempurl%
\url{https://api.semanticscholar.org/CorpusID:250175050}
\showURL{%
\tempurl}


\bibitem[You et~al\mbox{.}(2010)]%
        {you2010finger}
\bibfield{author}{\bibinfo{person}{Kyung-Jin You}, \bibinfo{person}{Ki-Won Rhee}, {and} \bibinfo{person}{Hyun-Chool Shin}.} \bibinfo{year}{2010}\natexlab{}.
\newblock \showarticletitle{Finger motion decoding using EMG signals corresponding various arm postures}.
\newblock \bibinfo{journal}{\emph{Experimental neurobiology}} \bibinfo{volume}{19}, \bibinfo{number}{1} (\bibinfo{year}{2010}), \bibinfo{pages}{54}.
\newblock


\bibitem[Yu et~al\mbox{.}(2024)]%
        {yu2024ring}
\bibfield{author}{\bibinfo{person}{Tianhong~Catherine Yu}, \bibinfo{person}{Guilin Hu}, \bibinfo{person}{Ruidong Zhang}, \bibinfo{person}{Hyunchul Lim}, \bibinfo{person}{Saif Mahmud}, \bibinfo{person}{Chi-Jung Lee}, \bibinfo{person}{Ke Li}, \bibinfo{person}{Devansh Agarwal}, \bibinfo{person}{Shuyang Nie}, \bibinfo{person}{Jinseok Oh}, {et~al\mbox{.}}} \bibinfo{year}{2024}\natexlab{}.
\newblock \showarticletitle{Ring-a-Pose: A Ring for Continuous Hand Pose Tracking}.
\newblock \bibinfo{journal}{\emph{Proceedings of the ACM on Interactive, Mobile, Wearable and Ubiquitous Technologies}} \bibinfo{volume}{8}, \bibinfo{number}{4} (\bibinfo{year}{2024}), \bibinfo{pages}{1--30}.
\newblock


\bibitem[Yuan et~al\mbox{.}(2017)]%
        {yuan2017bighand2}
\bibfield{author}{\bibinfo{person}{Shanxin Yuan}, \bibinfo{person}{Qi Ye}, \bibinfo{person}{Bjorn Stenger}, \bibinfo{person}{Siddhant Jain}, {and} \bibinfo{person}{Tae-Kyun Kim}.} \bibinfo{year}{2017}\natexlab{}.
\newblock \showarticletitle{Bighand2. 2m benchmark: Hand pose dataset and state of the art analysis}. In \bibinfo{booktitle}{\emph{Proceedings of the IEEE conference on computer vision and pattern recognition}}. \bibinfo{pages}{4866--4874}.
\newblock


\bibitem[Zatsiorsky et~al\mbox{.}(2000)]%
        {zatsiorsky2000enslaving}
\bibfield{author}{\bibinfo{person}{Vladimir~M Zatsiorsky}, \bibinfo{person}{Zong-Ming Li}, {and} \bibinfo{person}{Mark~L Latash}.} \bibinfo{year}{2000}\natexlab{}.
\newblock \showarticletitle{Enslaving effects in multi-finger force production}.
\newblock \bibinfo{journal}{\emph{Experimental brain research}}  \bibinfo{volume}{131} (\bibinfo{year}{2000}), \bibinfo{pages}{187--195}.
\newblock


\bibitem[Zhang et~al\mbox{.}(2016)]%
        {zhang20163d}
\bibfield{author}{\bibinfo{person}{Jiawei Zhang}, \bibinfo{person}{Jianbo Jiao}, \bibinfo{person}{Mingliang Chen}, \bibinfo{person}{Liangqiong Qu}, \bibinfo{person}{Xiaobin Xu}, {and} \bibinfo{person}{Qingxiong Yang}.} \bibinfo{year}{2016}\natexlab{}.
\newblock \showarticletitle{3d hand pose tracking and estimation using stereo matching}.
\newblock \bibinfo{journal}{\emph{arXiv preprint arXiv:1610.07214}} (\bibinfo{year}{2016}).
\newblock


\bibitem[Zheng et~al\mbox{.}(2024)]%
        {zheng2024prediction}
\bibfield{author}{\bibinfo{person}{Bofang Zheng}, \bibinfo{person}{Yixin Li}, \bibinfo{person}{Guanghua Xu}, \bibinfo{person}{Gang Wang}, {and} \bibinfo{person}{Yang Zheng}.} \bibinfo{year}{2024}\natexlab{}.
\newblock \showarticletitle{Prediction of Dexterous Finger Forces with Forearm Rotation using Motoneuron Discharges}.
\newblock \bibinfo{journal}{\emph{IEEE Transactions on Neural Systems and Rehabilitation Engineering}} (\bibinfo{year}{2024}).
\newblock


\bibitem[ZHOU(2022)]%
        {ZHOU2022LearningOT}
\bibfield{author}{\bibinfo{person}{HAO ZHOU}.} \bibinfo{year}{2022}\natexlab{}.
\newblock \showarticletitle{Learning on the Rings}.
\newblock \bibinfo{journal}{\emph{Proceedings of the ACM on Interactive, Mobile, Wearable and Ubiquitous Technologies}}  \bibinfo{volume}{6} (\bibinfo{year}{2022}), \bibinfo{pages}{1 -- 31}.
\newblock
\urldef\tempurl%
\url{https://api.semanticscholar.org/CorpusID:250055792}
\showURL{%
\tempurl}


\bibitem[Zhou et~al\mbox{.}(2022)]%
        {zhou2022learning}
\bibfield{author}{\bibinfo{person}{Hao Zhou}, \bibinfo{person}{Taiting Lu}, \bibinfo{person}{Yilin Liu}, \bibinfo{person}{Shijia Zhang}, {and} \bibinfo{person}{Mahanth Gowda}.} \bibinfo{year}{2022}\natexlab{}.
\newblock \showarticletitle{Learning on the rings: Self-supervised 3d finger motion tracking using wearable sensors}.
\newblock \bibinfo{journal}{\emph{Proceedings of the ACM on Interactive, Mobile, Wearable and Ubiquitous Technologies}} \bibinfo{volume}{6}, \bibinfo{number}{2} (\bibinfo{year}{2022}), \bibinfo{pages}{1--31}.
\newblock


\bibitem[Zhou et~al\mbox{.}(2023)]%
        {zhou2023one}
\bibfield{author}{\bibinfo{person}{Hao Zhou}, \bibinfo{person}{Taiting Lu}, \bibinfo{person}{Yilin Liu}, \bibinfo{person}{Shijia Zhang}, \bibinfo{person}{Runze Liu}, {and} \bibinfo{person}{Mahanth Gowda}.} \bibinfo{year}{2023}\natexlab{}.
\newblock \showarticletitle{One Ring to Rule Them All: An Open Source Smartring Platform for Finger Motion Analytics and Healthcare Applications}. In \bibinfo{booktitle}{\emph{Proceedings of the 8th ACM/IEEE Conference on Internet of Things Design and Implementation (IoTDI '23)}}. \bibinfo{publisher}{Association for Computing Machinery}, \bibinfo{address}{New York, NY, USA}, \bibinfo{pages}{27--38}.
\newblock
\urldef\tempurl%
\url{https://doi.org/10.1145/3576842.3582382}
\showDOI{\tempurl}


\end{thebibliography}

\newpage
\appendix












\end{document}